\documentclass[a4paper,12pt]{article}
	\usepackage{tikz}
	\usepackage{tabularx,ragged2e}
	\usepackage{pythontex} 
	\usetikzlibrary{patterns,perspective}
	\usepackage{pgfplots}
	   \usetikzlibrary{matrix,decorations.pathreplacing, calc, positioning,fit}
	\usetikzlibrary{positioning}
	\usetikzlibrary{calc}
	\usetikzlibrary{arrows,shapes,backgrounds}
	\usepackage{bbding}
	\usepackage{bbm}
	\usepackage{amsmath}
	\usepackage{amsthm}
	\usepackage{listings}
	\usepackage[margin=0cm]{caption}

	\usepackage{bm}
	\usepackage{bbold}
	\usepackage{appendix}
	\usepackage{graphicx}
	\usepackage{natbib}
	\usepackage{listings}
	\usepackage[ruled,vlined]{algorithm2e}
	\usepackage{hyperref}
	\usepackage{cleveref}
	\usepackage{enumitem}
	\usepackage{longtable}

	\DeclareMathAlphabet{\pazocal}{OMS}{zplm}{m}{n}

	\newcommand{\bbeta}{{\boldsymbol{\beta}}}

	\newcommand{\bX}{\mathbf{X}}

	\usepackage[utf8]{inputenc}
	
	\DeclareFixedFont{\ttb}{T1}{txtt}{bx}{n}{12} 
	\DeclareFixedFont{\ttm}{T1}{txtt}{m}{n}{12}  
	
	\usepackage{color}
	\definecolor{deepblue}{rgb}{0,0,0.5}
	\definecolor{deepred}{rgb}{0.6,0,0}
	\definecolor{deepgreen}{rgb}{0,0.5,0}
	
	\usepackage{listings}
	
	\newcommand\pythonstyle{\lstset{
			language=Python,
			basicstyle=\ttm,
			morekeywords={self},              
			keywordstyle=\ttb\color{deepblue},
			emph={MyClass,__init__},          
			emphstyle=\ttb\color{deepred},    
			stringstyle=\color{deepgreen},
			frame=tb,                         
			showstringspaces=false
	}}

	\lstnewenvironment{python}[1][basicstyle=\small]
	{
		\pythonstyle
		\lstset{#1}
	}
	{}
	
	\newcommand\pythoninline[1]{{\pythonstyle\lstinline!#1!}}
	
\begin{document}
	\title{Changes in Risk Appreciation, and Short Memory of House Buyers When the Market is Hot, a Case Study of Christchurch, New Zealand}
	\author{\begin{tabular}{ccc}
	Emil Mendoza\footnote{School of Mathematics and Statistics, University of Canterbury, Private Bag 4800, Christchurch 8140, New Zealand, emil.mendoza@pg.canterbury.ac.nz} \footnote{Corresponding author} &
	Fabian Dunker\footnote{School of Mathematics and Statistics, University of Canterbury, Private Bag 4800, Christchurch 8140, New Zealand, fabian.dunker@canterbury.ac.nz} & Marco Reale\footnote{School of Mathematics and Statistics, University of Canterbury, Private Bag 4800, Christchurch 8140, New Zealand, marco.reale@canterbury.ac.nz}\\
	\small{University of Canterbury}  & \small{University of Canterbury}  & \small{University of Canterbury}
	\end{tabular}
	}
	\maketitle
	
	\begin{abstract}
		\noindent 
		In this paper house prices in Christchurch are analyzed over three distinct periods of time: post-2011 earthquake, pre-COVID-19 lockdown, and post-COVID-19 lockdown using the well-established hedonic price model. 
Results show that buyers, in periods that are temporally distant from the 2011 Christchurch earthquake, value the risk of potential earthquake damage to a property differently from buyers soon after the earthquake. We find that there are observable shifts in hedonic prices across the different time periods, specifically for section size pre and post COVID-19 lockdown.
	\end{abstract}

	\section{Introduction}\label{sec:intro}
	
	Many property markets around the world post COVID-19 lockdowns were dominated by significant price inflation due to low interest rates. However, there is preliminary evidence of more complex changes than just an increase in prices. For example, some buyers seem to appreciate larger sections after having been in lockdown. In this paper we analyze hedonic prices and marginal willingness to pay of house buyers for different features of a property using the hedonic house price model. 
We compare the property market over three periods. Relatively stable periods during 2014 to 2016, the period pre COVID-19, and a hot market post COVID-19. We find that hedonic prices change and buyers seem to have a different attitude towards risk associated with the location of the house. 
	
	One of our main findings is that in the period immediately following the 2011 earthquake, buyers were willing to pay more for locations that were known to have a relatively low risk of sustaining damage due to soil liquefaction in the event of another major earthquake. In the other periods, 9 to 10 years after the major Christchurch earthquake, the effects of this additional risk to the property was not as significant. We found that, post-lockdown, buyers were willing to pay more for a larger section size, compared to the pre-lockdown period. 
	
	The hedonic house price model is a very popular tool and has been widely used in the literature. It builds on hedonic price models for differentiated products introduced by \cite{lancaster1966new}. These models explain product prices as a function of observable product characteristics. \cite{rosen1974hedonic} connects the model to consumer demand in an equilibrium market which allows to infer the consumer's marginal willingness to pay for each product characteristic from observations of prices. 
	The main assumptions necessary for this deduction are a market equilibrium, i.e. no consumer can increase their utility by choosing a different product. In addition, consumers are fully informed, free to choose any other product and to purchase a continuous level of each characteristic. For detailed discussions of these assumptions in the context of property markets see \cite{Palmquist:05}.
	
	
	
	A large number of studies did use hedonic price
	theory to analyze the property market. We can just mention a few. Heteroskedasticity in the hedonic
	price model induced by dwelling age is explored in \cite{goodman1995age} which describes a vintage effect that drives up a house’s price. An
	extension is \cite{fletcher2000heteroscedasticity}
	which identifies other possible sources of heteroskedasticity, specifically the
	area of the property.
	Another important work using the hedonic price model is spatial autocorrelation and spatial heterogeneity. In \cite{dubin1992spatial}, using data
	from Baltimore, a new approach to evaluating the strength of the effect of
	spatially related variables such as neighbourhood quality and accessibility is
	implemented. The method builds on the hypothesis that the spatial auto-
	correlation in the error term will reveal the effect of neighbourhood quality
	and accessibility when they are left out as initial explanatory variables.
	In \cite{helbich2014spatial}, the importance of spatial heterogeneity
	is demonstrated in the case of Austria, where it was shown that effects of
	variables can be overestimated if spatial effects are not considered.
	The hedonic house price model has also been used to examine the effects
	of specific characteristics of a house on its price. The effect of transportation
	infrastructure on house prices in Sydney, Australia was examined in \cite{lieske2021novel}. Effects of architectural features were analyzed with a spatial differences in differences method by \cite{AH:18}. The effect of upzoning on house prices in Auckland, New
	Zealand was examined \cite{greenaway2021effect}. And, the effects
	of school zones and proximity to cellular towers on house prices in New
	Zealand was examined in \cite{rehm2008impact}, and \cite{filippova2011impact}.
	In \cite{limsombunchai2004house} the hedonic
	model was compared to an Artificial Neural Network in terms of
	predictive capability based on Christchurch data. Finally, the impact of earthquake risk on property prices in Japan has been analyzed in \cite{ikefuji2022earthquake}.
	
	The outline of the paper is as follows. Section \ref{sec:data} gives an overview about the data sources and quality. The model is explained in Section \ref{sec:method}. Section \ref{sec:results} discusses different findings. The paper closes with in conclusion in Section \ref{sec:conclusion}. The Appendix contains further discussions of spatial and quarterly dummy variables.
	\section{Data}\label{sec:data}
	
	The data consists of log real house prices and a selection of structural, neighborhood, and locational characteristics. The data were obtained from variety of different sources. Data such as nominal sales prices, dates of sale, structural characteristics, and zoning were obtained through TradeMe and QV. Real house prices are obtained by dividing the nominal house prices by the House Price Index (HPI) obtained from the Reserve Bank of New Zealand. Geospatial related characteristics such as distance to schools, and the CBD were obtained by geocoding the addresses through the Googlemaps API and calculating the Haversine distance between two pairs of longitude and latitudes. 
	
	Another very important geospatial characteristic is the technical category (TC) of the land on which a property is built on. The technical category is a classification of land that indicates the likelihood of a property to experience soil-liquefaction related damage in the event of an earthquake. TC1 zones are the least likely to sustain liquefaction related damage, and TC3 zones face the possibility of significant damage related to liquefaction in the future event of a significant earthquake, with TC2 zones in between. These data were obtained from Canterbury Maps Open Data. 
	
	The 2013 and 2018 New Zealand Deprivation Indices, which are comprised of variety of measures of deprivation such as income, education, and housing quality, \cite{atkinson2021new}, were used as features that represent the overall condition of a neighbourhood. The index was used as a proxy variable for how good the area is to live in. The deprivation index is available at a meshblock level which is the smallest geographical unit for which statistical information is collected, see \cite{atkinson2021new}.
	
	The 2010 New Zealand school zones were obtained from Koordinates, which is an open access database for geographic data, as a geographical comma delimited file. The geocoded properties were then assigned using their latitude and longitudes by determining if it lies within the polygonal area of a specific school zone. In 2019, new demarcations for school zones were issued and initially made known. These changes have not taken effect as of 2021. The data is also not yet available in an open access format. However, the changes may have already affected property prices. We kept this in mind when comparing the effects of school zones across different time frames. 
	
	The data was cleaned and filtered before it was used for model fitting. Three subsets were constructed based on different time periods, 01-Jan-2014 to 31-Dec-2015, 01-March-2019 to 29-Feb-2020, and 01-May-2020 to 30-Jun-2021. The first time period corresponds to a time frame that is close to the 2011 earthquake that caused significant damage to a large number of properties in  Christchurch, but is a period where the property market returned to stable conditions. The second time period corresponds to a time frame that is temporally distant from the 2011 earthquake and before the 2020 nationwide lockdown due to COVID-19. The final time period corresponds to a time period after the easing of lockdown restrictions and when the market was turning hot. 
	
	To ensure that there were no erroneous entries and significant outliers, sensible filters were applied. Properties whose distance to the CBD was greater than fifteen kilometers were removed to ensure only properties within Christchurch were considered. Properties with floor areas less than $60m^2$, land areas less than $200 m^2$, and land areas greater than $3000m^2$ were removed. 
	
	Furthermore, only 'Residential Suburban' properties were used. This was done in order to control for the unobserved effects attached to different zoning codes, as it carries with it different rules for land use. Properties used in the analysis were also limited to those that have assigned technical categories. The summary of features used as input data for the pre-lockdown period can be seen in Table \ref{tab:input_data_pre}, the additional tables for other time periods can be found in the Appendix. 
	
	Finally, there is an important consideration to make when drawing conclusions derived from the subset of data spanning 2014 to 2016. Due to constraints and limitations in the process of obtaining the data, we were limited to using properties that were sold both in the period 2019 to 2021 and 2014 to 2016. This leads to some form of a selection bias, since there might be particular qualities that are associated with properties that are resold at a higher frequency. However, the effect of this is not within the scope of the study, and we assume the case that the selection bias has minimal impact on the specific analysis done in this paper.
	
	\begin{table}[!h]
		\caption{Input data for pre-lockdown period $n=1091$}
		\centering 
		\hspace*{-2cm}\begin{tabular}{c lllll} 
			\hline\hline \\ [-1.5ex]
			\textbf{Variable} & \textbf{Data Type} & \textbf{Min} & \textbf{Max} & \textbf{Mean} & \textbf{Std. Dev.} \\\hline
			\\ [-1.5ex]
			\multicolumn{1}{p{3cm}}{\raggedright Log Real Price} & \multicolumn{1}{p{3cm}}{\raggedright Floating point} & \multicolumn{1}{p{1cm}}{\raggedright 4.719} & 
			\multicolumn{1}{p{1.5cm}}{\raggedright 7.588} & 
			\multicolumn{1}{p{1.5cm}}{\raggedright 5.298} & 
			\multicolumn{1}{p{1.5cm}}{\raggedright 0.337} \\ 
			\multicolumn{1}{p{3cm}}{\raggedright Bedrooms} & 
			\multicolumn{1}{p{3cm}}{\raggedright Integer} & 
			\multicolumn{1}{p{1.5cm}}{\raggedright 2} & 
			\multicolumn{1}{p{1.5cm}}{\raggedright 9} & 
			\multicolumn{1}{p{1.5cm}}{\raggedright -} & 
			\multicolumn{1}{p{1.5cm}}{\raggedright 0.794} \\
			\multicolumn{1}{p{3cm}}{\raggedright Bathooms} & 
			\multicolumn{1}{p{3cm}}{\raggedright Integer} & 
			\multicolumn{1}{p{1cm}}{\raggedright 1} & 
			\multicolumn{1}{p{1.5cm}}{\raggedright 4} & 
			\multicolumn{1}{p{1.5cm}}{\raggedright -} & 
			\multicolumn{1}{p{1.5cm}}{\raggedright 0.608} \\ 
			\multicolumn{1}{p{3cm}}{\raggedright Toilets} & 
			\multicolumn{1}{p{3cm}}{\raggedright Integer} & 
			\multicolumn{1}{p{1.5cm}}{\raggedright 1} & 
			\multicolumn{1}{p{1.5cm}}{\raggedright 4} & 
			\multicolumn{1}{p{1.5cm}}{\raggedright -} & 
			\multicolumn{1}{p{1.5cm}}{\raggedright 0.773} \\ 
			\multicolumn{1}{p{3cm}}{\raggedright Carparks} & 
			\multicolumn{1}{p{3cm}}{\raggedright Integer} & 
			\multicolumn{1}{p{1.5cm}}{\raggedright 1} & 
			\multicolumn{1}{p{1.5cm}}{\raggedright 6} & 
			\multicolumn{1}{p{1.5cm}}{\raggedright -} & 
			\multicolumn{1}{p{1.5cm}}{\raggedright 0.567} \\
			\multicolumn{1}{p{3cm}}{\raggedright Floor Area} & 
			\multicolumn{1}{p{3cm}}{\raggedright Floating Point} & 
			\multicolumn{1}{p{1.5cm}}{\raggedright 62.0$m^2$} & 
			\multicolumn{1}{p{1.5cm}}{\raggedright 498.0$m^2$} & 
			\multicolumn{1}{p{1.5cm}}{\raggedright 161.6$m^2$} & 
			\multicolumn{1}{p{1cm}}{\raggedright 64.031} \\ 
			\multicolumn{1}{p{3cm}}{\raggedright Land Area} & 
			\multicolumn{1}{p{3cm}}{\raggedright Floating Point} & 
			\multicolumn{1}{p{1.5cm}}{\raggedright 219.0$m^2$} & 
			\multicolumn{1}{p{1.5cm}}{\raggedright 2925.0$m^2$} & 
			\multicolumn{1}{p{1.5cm}}{\raggedright 690.8$m^2$} & 
			\multicolumn{1}{p{1.56cm}}{\raggedright 203.880} \\ 
			\multicolumn{1}{p{3cm}}{\raggedright Position of Property} & 
			\multicolumn{1}{p{3cm}}{\raggedright Categorical} & 
			\multicolumn{1}{p{1.5cm}}{\raggedright -} & 
			\multicolumn{1}{p{1.5cm}}{\raggedright -} & 
			\multicolumn{1}{p{1.5cm}}{\raggedright -} & 
			\multicolumn{1}{p{1.56cm}}{\raggedright -} \\ 
			\multicolumn{1}{p{3cm}}{\raggedright Roof Condition} & 
			\multicolumn{1}{p{3cm}}{\raggedright Categorical} & 
			\multicolumn{1}{p{1.5cm}}{\raggedright -} & 
			\multicolumn{1}{p{1.5cm}}{\raggedright -} & 
			\multicolumn{1}{p{1.5cm}}{\raggedright -} & 
			\multicolumn{1}{p{1.56cm}}{\raggedright -} \\ 
			\multicolumn{1}{p{3cm}}{\raggedright View from Property} & 
			\multicolumn{1}{p{3cm}}{\raggedright Categorical} & 
			\multicolumn{1}{p{1.5cm}}{\raggedright -} & 
			\multicolumn{1}{p{1.5cm}}{\raggedright -} & 
			\multicolumn{1}{p{1.5cm}}{\raggedright -} & 
			\multicolumn{1}{p{1.56cm}}{\raggedright -} \\
			\multicolumn{1}{p{3cm}}{\raggedright Wall Condition} & 
			\multicolumn{1}{p{3cm}}{\raggedright Categorical} & 
			\multicolumn{1}{p{1.5cm}}{\raggedright -} & 
			\multicolumn{1}{p{1.5cm}}{\raggedright -} & 
			\multicolumn{1}{p{1.5cm}}{\raggedright -} & 
			\multicolumn{1}{p{1.56cm}}{\raggedright -} \\ 
			\multicolumn{1}{p{3cm}}{\raggedright Dist. CBD} & 
			\multicolumn{1}{p{3cm}}{\raggedright Floating point} & 
			\multicolumn{1}{p{1.5cm}}{\raggedright 0.237$km$} & 
			\multicolumn{1}{p{1.5cm}}{\raggedright 13.498$km$} & 
			\multicolumn{1}{p{1.5cm}}{\raggedright 6.018$km$} & 
			\multicolumn{1}{p{1.56cm}}{\raggedright 1.927} \\ 
			\multicolumn{1}{p{3cm}}{\raggedright Dist. to Secondary} & 
			\multicolumn{1}{p{3cm}}{\raggedright Floating point} & 
			\multicolumn{1}{p{1.5cm}}{\raggedright 0.101$km$} & 
			\multicolumn{1}{p{1.5cm}}{\raggedright 8.841$km$} & 
			\multicolumn{1}{p{1.5cm}}{\raggedright 1.760$km$} & 
			\multicolumn{1}{p{1.56cm}}{\raggedright 1.213} \\
			\multicolumn{1}{p{3cm}}{\raggedright Dist. to Secondary} & 
			\multicolumn{1}{p{3cm}}{\raggedright Floating point} & 
			\multicolumn{1}{p{1.5cm}}{\raggedright 0.101$km$} & 
			\multicolumn{1}{p{1.5cm}}{\raggedright 8.841$km$} & 
			\multicolumn{1}{p{1.5cm}}{\raggedright 1.760$km$} & 
			\multicolumn{1}{p{1.56cm}}{\raggedright 1.213} \\  
			\multicolumn{1}{p{3cm}}{\raggedright Technical Category} & 
			\multicolumn{1}{p{3cm}}{\raggedright Categorical} & 
			\multicolumn{1}{p{1.5cm}}{\raggedright -} & 
			\multicolumn{1}{p{1.5cm}}{\raggedright -} & 
			\multicolumn{1}{p{1.5cm}}{\raggedright -} & 
			\multicolumn{1}{p{1.56cm}}{\raggedright -} \\  
			\multicolumn{1}{p{3cm}}{\raggedright Age Purchased} & 
			\multicolumn{1}{p{3cm}}{\raggedright Integer} & 
			\multicolumn{1}{p{1.5cm}}{\raggedright 0} & 
			\multicolumn{1}{p{1.5cm}}{\raggedright 119} & 
			\multicolumn{1}{p{1.5cm}}{\raggedright 47.67} & 
			\multicolumn{1}{p{1.56cm}}{\raggedright 25.884} \\ 
			\multicolumn{1}{p{3cm}}{\raggedright NZ Deprivation Score} & 
			\multicolumn{1}{p{3cm}}{\raggedright Integer} & 
			\multicolumn{1}{p{1.5cm}}{\raggedright 846} & 
			\multicolumn{1}{p{1.5cm}}{\raggedright 1249} & 
			\multicolumn{1}{p{1.5cm}}{\raggedright 969.9} & 
			\multicolumn{1}{p{1.56cm}}{\raggedright 70.870} \\  
			\multicolumn{1}{p{3cm}}{\raggedright School Zone} & 
			\multicolumn{1}{p{3cm}}{\raggedright Categorical} & 
			\multicolumn{1}{p{1.5cm}}{\raggedright -} & 
			\multicolumn{1}{p{1.5cm}}{\raggedright -} & 
			\multicolumn{1}{p{1.5cm}}{\raggedright -} & 
			\multicolumn{1}{p{1.56cm}}{\raggedright -} \\ 
			\multicolumn{1}{p{3cm}}{\raggedright Ward} & 
			\multicolumn{1}{p{1.5cm}}{\raggedright Categorical} & 
			\multicolumn{1}{p{1.5cm}}{\raggedright -} & 
			\multicolumn{1}{p{1.5cm}}{\raggedright -} & 
			\multicolumn{1}{p{1.56cm}}{\raggedright -} \\ 
	 		\multicolumn{1}{p{4cm}}{\raggedright (Age Purchased)$^2$} & 
			\multicolumn{1}{p{1.5cm}}{\raggedright Integer} & 
			\multicolumn{1}{p{1.5cm}}{\raggedright 0} & 
			\multicolumn{1}{p{1.5cm}}{\raggedright 14161} & 
			\multicolumn{1}{p{1.5cm}}{\raggedright 2941.890} &
			\multicolumn{1}{p{1.5cm}}{\raggedright 2722.924} \\ 
					\hline
		\end{tabular}\hspace*{-2cm}
		\label{tab:input_data_pre}
	\end{table}
	
	\section{Method} \label{sec:method}
	
	The first model considered is the standard multiple linear regression model given by the equation below,
	
	\begin{equation}
	\ln P  = \bbeta\bX + \boldmath{\epsilon}
	\label{eqn:hpm_base}
	\end{equation}
	
	where $\bX$ is a matrix of regressors listed in Table \ref{tab:input_data_pre}, $\bbeta$ is a vector of slopes and an intercept, $\boldmath{\epsilon}$ is the error term, and $\boldmath{P}$ are the real property prices.
	
An alternative approach is to use nominal instead of deflated log-prices as depended variable and account for price inflation by adding time dummies to the regressors, see \cite{Triplett04}. We present the results of the related time dummy model in Appendix \ref{app:timedummy}. As one would expect, the results are similar to using real prices.
	
	An important consideration in fitting the hedonic model is omitted-variable bias that manifests as spatial correlation, \cite{bishop2020best}. One variable of interest, technical category (TC) is an inherently spatial characteristic of a residential property. Great care was taken in specifying the model to prevent misattributing variation in the response variable to TC rather than other spatial characteristics. However, it is unlikely that all attributes of a property that are geographically determined can be accounted for. 
	
	An approach suggested by \cite{bishop2020best} to address the problem of omitted-variable bias is to assign a dummy variable that determines the geographical location of a property such as as a zip code, or a suburb. We follow the methodology in \cite{gibson2007house} where the wards in which properties are in were included as a variable. 

	We have investigated the use of suburbs as the geographical feature mainly used to capture additional variation in the response due to geographical location. While the use of suburbs does better specify the model indicated by an increase in the model’s R-squared it also introduces the problem of collinearity with the TC variable, and hence masks its effect. More specifically, in the 2020 to 2021 period subset of data, all TC1 classified properties are found in only thirteen out of the sixty-five suburbs, and out of all the properties located in these suburbs included in the data set 74\% are TC1. The two other subsets of data for the 2014 to 2016, and 2019 to 2020 periods have similar statistics, as is shown in Table \ref{tab:contigency} in Appendix \ref{app:additional}. Given that this masks the effect of the variable that we are interested in, we opted to retain the usage of wards, distance to the city center and secondary schools, and the deprivation index as the controls to account for variation in the response due to geographical location. A summary of the models, and the relevant contingency table is available in Appendix \ref{app:additional}.

	It was made apparent in \cite{mcclay2003impact}, \cite{gibson2007house}, and \cite{rehm2008impact} that school zones are a significant driver of house prices in New Zealand. In our analysis, Burnside, Riccarton, Christchurch Girls', and Christchurch Boys' school zones were considered as in \cite{gibson2007house}, and additionally we also included the Cashmere Highschool zone. These school zones were used as variables to account for the premium commanded by being located in these desirable school zones.
	
	Ideally other environmental amenities would be included such as air quality and water quality   as geographical features; however, these data are unavailable at the desired level of granularity.

	\section{Results and Discussion}\label{sec:results}
	
	The results for the hedonic regression models are presented in Table \ref{tab:regression_results}. All the models for the periods 2014 to 2016, 2019 to 2020 (pre-lockdown), and 2020 to 2021 (post-lockdown) were significant at a 0.001 level. The models are all well-performing as each had an $R$-squared of 0.751, 0.808, and 0.823 respectively.
	
	We focus our investigation on shifts in buyers' marginal willingness to pay across the different time periods for certain structural attributes, changes in the assessment of risk of sustaining earthquake damage brought about by a property's location. 
	Locational and neighborhood attributes such as school zones, deprivation index, and wards primarily serve as control variables to prevent misattribution of variation to the locational variable of interest, technical category (TC). In-depth discussion on the effects of these control variables across the different models can be found in Appendix \ref{app:discussion}.
	
	Analysis of the effect of school zoning on house prices in Christchurch has been done in \cite{mcclay2003impact}, and \cite{gibson2007house}. The results that we obtained generally agree with the findings in those studies. We assume that the results in these studies are more precise than our results on school zones, because the models used in  \cite{mcclay2003impact} and \cite{gibson2007house} are specifically designed for analyzing school zones, while our focus is on the effect of the technical categories.
	

When interpreting the results it needs to be noted that changes in implicit prices can have several sources. One source can be a change in buyers' valuation of the attributes. A second effect can come from a shift in supply or demand which leads to a new market equilibrium. The first two periods in our study saw a relatively stable property market in Christchurch with almost flat price increases. We assume that the market equilibrium did not shift much between these two periods. The third period was characterized by decreased interest rates, increased demand, and rising prices. We assume that changes of hedonic prices in this period are driven by both, the new situation in the market, and a change of buyers' valuation of certain attributes after going through the experience of a lockdown.
	
	\subsection*{Structural Attributes}
	
	Structural attributes such as the number of bedrooms, bathrooms, toilets, and carparks were generally insignificant for the periods 2014 to 2016 and pre-lockdown, while the number of bedrooms, and the number of toilets are significant at 0.001 and 0.01 levels respectively for the post-lockdown period. The insignificance of most of these structural attributes in the two periods is understandable as they are highly correlated with each other, and also with floor area and land area.
	
	At first glance, it seems peculiar that the number of bedrooms has a negative effect on the price of a property with all else held equal, leading to approximately a 4.5\% decrease in a property's price for each additional bedroom unit. However, it is important to consider the context in which this  interpretation is made i.e. all other parameters are held constant. With this in mind, a plausible explanation is that the number of bedrooms could be less important than the size of these bedrooms from a buyer's perspective. If floor area is held constant, an increase in the number of bedrooms limits the area of each bedroom and likewise other livable parts of the house, which the data and model tell us is not desirable. 
	 
	The floor area and land area of a property were significant for all the models up to a 0.001 level, with the sole exception of land area being significant only at a 0.1 level for the 2014 to 2016 period. The effects of these attributes across the three models are all positive as expected.
	
	An interesting observation is that buyers' willingness to pay for floor area and land area seems to be increasing over time. It is even more interesting to look at the difference in willingness to pay for land area between the pre-lockdown and post-lockdown periods which had a bigger change compared to the difference in the effect of floor area. It seems to suggest that buyers, post-lockdown, prefer a much larger section size than they did pre-lockdown. This might be explained by the fact that a larger section size allows for more freedoms in the event of another strict lockdown. 
	
	\setlength\LTleft{0cm}
	\setlength\LTright{0pt plus 1fill minus 1fill}%
	\small
	\setlength\LTleft{-2cm}
	\setlength\LTright{0pt plus 1fill minus 1fill}
	\begin{longtable}{c lll} 
		\caption{Hedonic regression estimates by time period}
		\label{tab:regression_results} \\
		\hline\hline \\ [-1.5ex]
		\textbf{Variable} & \textbf{2014-2016} & \textbf{2019-2020} & \textbf{2020-2021} \\
		& Coefficient (SE) & Coefficient (SE) &  Coefficient (SE) \\
		\hline
		\\ [-1.5ex]
		\multicolumn{1}{p{4cm}}{\raggedright Intercept} & 
		\multicolumn{1}{p{4cm}}{\raggedright 6.008*** (2.161e-1)} & 
		\multicolumn{1}{p{4cm}}{\raggedright 5.572*** (9.888e-2)} & 
		\multicolumn{1}{p{4cm}}{\raggedright 5.622*** (8.017e-2)} \\ 
		\multicolumn{1}{p{4cm}}{\raggedright Bedrooms} & 
		\multicolumn{1}{p{4cm}}{\raggedright 2.618e- (2.950e-2)} & 
		\multicolumn{1}{p{4cm}}{\raggedright -3.748e-4 (1.129e-2) } & 
		\multicolumn{1}{p{4cm}}{\raggedright -4.603e-2*** (9.502e-3)} \\
		\multicolumn{1}{p{4cm}}{\raggedright Bathrooms} & 
		\multicolumn{1}{p{4cm}}{\raggedright -2.615e-2 (3.580e-2)} & 
		\multicolumn{1}{p{4cm}}{\raggedright 2.963e-3 (1.551e-2)} & 
		\multicolumn{1}{p{4cm}}{\raggedright  2.236e-2$\dagger$ (1.247e-2)} \\ 
		\multicolumn{1}{p{4cm}}{\raggedright Toilets} & 
		\multicolumn{1}{p{4cm}}{\raggedright 4.900e-2 (3.261e-2)} & 
		\multicolumn{1}{p{4cm}}{\raggedright  1.782e-2  (1.331e-2)} & 
		\multicolumn{1}{p{4cm}}{\raggedright 2.808e-2 ** (1.033e-2)} \\ 
		\multicolumn{1}{p{4cm}}{\raggedright Carparks} & 
		\multicolumn{1}{p{4cm}}{\raggedright 2.227e-2  (1.895e-2)} & 
		\multicolumn{1}{p{4cm}}{\raggedright 2.790e-3 (9.303e-3)} & 
		\multicolumn{1}{p{4cm}}{\raggedright 1.616e-3 (6.807e-3)} \\
		\multicolumn{1}{p{4cm}}{\raggedright Floor Area} & 
		\multicolumn{1}{p{4cm}}{\raggedright 1.876e-3***(3.674e-4) } & 
		\multicolumn{1}{p{4cm}}{\raggedright 2.202e-3*** (1.564e-4)} & 
		\multicolumn{1}{p{4cm}}{\raggedright  2.349e-3*** (1.238e-4)} \\ 
		\multicolumn{1}{p{4cm}}{\raggedright Land Area} & 
		\multicolumn{1}{p{4cm}}{\raggedright 1.068e-4$\dagger$ (6.470e-5)} & 
		\multicolumn{1}{p{4cm}}{\raggedright 2.110e-4***(2.568e-5)} & 
		\multicolumn{1}{p{4cm}}{\raggedright 2.488e-4*** (2.113e-5)} \\ 
		\multicolumn{1}{p{4cm}}{\raggedright RoofCondition Fair} & 
		\multicolumn{1}{p{4cm}}{\raggedright 1.559e-2 (2.292e-1)} & 
		\multicolumn{1}{p{4cm}}{\raggedright 2.704e-2 (7.450e-2) } & 
		\multicolumn{1}{p{4cm}}{\raggedright -4.566e-2 (6.329e-2) } \\
		\multicolumn{1}{p{4cm}}{\raggedright RoofCondition Good} & 
		\multicolumn{1}{p{4cm}}{\raggedright 3.587e-3 (4.956e-2 ) } & 
		\multicolumn{1}{p{4cm}}{\raggedright -4.104e-3 (9.616e-3) } & 
		\multicolumn{1}{p{4cm}}{\raggedright  9.272e-3 (1.870e-2)} \\ 
		\multicolumn{1}{p{4cm}}{\raggedright RoofCondition Mixed} & 
		\multicolumn{1}{p{4cm}}{\raggedright -4.804e-1*** (9.930e-2)} & 
		\multicolumn{1}{p{4cm}}{\raggedright -1.090e-1** (3.790e-2) } & 
		\multicolumn{1}{p{4cm}}{\raggedright -1.843e-1*** (3.974e-2)} \\
		\multicolumn{1}{p{4cm}}{\raggedright RoofCondition Poor} & 
		\multicolumn{1}{p{4cm}}{\raggedright -} & 
		\multicolumn{1}{p{4cm}}{\raggedright -1.812e-1 (1.543e-1)} & 
		\multicolumn{1}{p{4cm}}{\raggedright 2.334e-2 (1.474e-1)} \\ 
		\multicolumn{1}{p{4cm}}{\raggedright View - Has view } & 
		\multicolumn{1}{p{4cm}}{\raggedright 1.037e-1 (6.841e-2) } & 
		\multicolumn{1}{p{4cm}}{\raggedright 2.768e-2 (2.552e-2)} & 
		\multicolumn{1}{p{4cm}}{\raggedright 4.307e-2$\dagger$ (2.235e-2)} \\
		\multicolumn{1}{p{4cm}}{\raggedright WallCondition Fair} & 
		\multicolumn{1}{p{4cm}}{\raggedright -2.271e-2 (1.881e-1) } & 
		\multicolumn{1}{p{4cm}}{\raggedright -6.387e-2 (6.601e-2 )} & 
		\multicolumn{1}{p{4cm}}{\raggedright -2.066e-2 (5.095e-2) } \\
		\multicolumn{1}{p{4cm}}{\raggedright WallCondition Good} & 
		\multicolumn{1}{p{4cm}}{\raggedright 2.197e-2 (4.873e-2)} & 
		\multicolumn{1}{p{4cm}}{\raggedright  7.225e-3 (2.360e-2} & 
		\multicolumn{1}{p{4cm}}{\raggedright 2.401e-2 (1.864e-2)} \\ 
		\multicolumn{1}{p{4cm}}{\raggedright WallCondition Mixed} & 
		\multicolumn{1}{p{4cm}}{\raggedright NA} & 
		\multicolumn{1}{p{4cm}}{\raggedright NA} & 
		\multicolumn{1}{p{4cm}}{\raggedright NA} \\ 
		\multicolumn{1}{p{4cm}}{\raggedright WallCondition Poor} & 
		\multicolumn{1}{p{4cm}}{\raggedright -} & 
		\multicolumn{1}{p{4cm}}{\raggedright -5.496e-2 (1.537e-1) } & 
		\multicolumn{1}{p{4cm}}{\raggedright NA} \\ 
		\multicolumn{1}{p{4cm}}{\raggedright Dist. CBD} & 
		\multicolumn{1}{p{4cm}}{\raggedright -4.302e-2*** (9.835e-3)} & 
		\multicolumn{1}{p{4cm}}{\raggedright -1.597e-2*** (4.585e-3 )} & 
		\multicolumn{1}{p{4cm}}{\raggedright -2.910e-2*** (3.697e-3)} \\ 
		\multicolumn{1}{p{4cm}}{\raggedright Dist. to Secondary} & 
		\multicolumn{1}{p{4cm}}{\raggedright 6.753e-2*** (1.248e-2)} & 
		\multicolumn{1}{p{4cm}}{\raggedright 2.738e-2*** (5.788e-3)} & 
		\multicolumn{1}{p{4cm}}{\raggedright 4.342e-2*** (4.847e-3)} \\  
		\multicolumn{1}{p{4cm}}{\raggedright TC2} & 
		\multicolumn{1}{p{4cm}}{\raggedright -1.272e-1** (4.792e-2)} & 
		\multicolumn{1}{p{4cm}}{\raggedright -2.129e-2 (2.102e-2)} & 
		\multicolumn{1}{p{4cm}}{\raggedright -5.473e-2*** (1.659e-2)} \\ 
		\multicolumn{1}{p{4cm}}{\raggedright TC3} & 
		\multicolumn{1}{p{4cm}}{\raggedright -2.474e-1*** (5.107e-2)} & 
		\multicolumn{1}{p{4cm}}{\raggedright -4.581e-2*(2.215e-2)} & 
		\multicolumn{1}{p{4cm}}{\raggedright -4.065e-2* (1.789e-2 )} \\   
		\multicolumn{1}{p{4cm}}{\raggedright Age Purchased} & 
		\multicolumn{1}{p{4cm}}{\raggedright -2.146e-3  (2.102e-3)} & 
		\multicolumn{1}{p{4cm}}{\raggedright -7.461e-3*** (8.331e-4)} & 
		\multicolumn{1}{p{4cm}}{\raggedright -4.969e-3*** (6.875e-4)} \\ 
		\multicolumn{1}{p{4cm}}{\raggedright (Age Purchased)$^2$} & 
		\multicolumn{1}{p{4cm}}{\raggedright 1.900e-5 (1.698e-5)} & 
		\multicolumn{1}{p{4cm}}{\raggedright 6.050e-5*** (7.117e-6)} & 
		\multicolumn{1}{p{4cm}}{\raggedright 4.010e-5*** (5.786e-6)} \\ [1ex]
		\multicolumn{1}{p{4cm}}{\raggedright NZ Deprivation Score} & 
		\multicolumn{1}{p{4cm}}{\raggedright -7.997e-4*** (1.818e-4)} & 
		\multicolumn{1}{p{4cm}}{\raggedright -6.447e-4*** (8.399e-5)} & 
		\multicolumn{1}{p{4cm}}{\raggedright -6.413e-4*** (6.813e-5)} \\  
		\multicolumn{1}{p{4cm}}{\raggedright Burnside} & 
		\multicolumn{1}{p{4cm}}{\raggedright -1.354e-3 (9.992e-2 )} & 
		\multicolumn{1}{p{4cm}}{\raggedright  1.085e-1** (4.094e-2)} & 
		\multicolumn{1}{p{4cm}}{\raggedright 2.100e-2 (3.449e-2)} \\
		\multicolumn{1}{p{4cm}}{\raggedright BurnsideCHCBoys} & 
		\multicolumn{1}{p{4cm}}{\raggedright 1.589e-1 (1.418e-1) } & 
		\multicolumn{1}{p{4cm}}{\raggedright 3.011e-1*** (5.826e-2)} & 
		\multicolumn{1}{p{4cm}}{\raggedright 2.269e-1*** (4.709e-2)} \\
		\multicolumn{1}{p{4cm}}{\raggedright BurnsideCHCBoysGirls} & 
		\multicolumn{1}{p{4cm}}{\raggedright -} & 
		\multicolumn{1}{p{4cm}}{\raggedright -} & 
		\multicolumn{1}{p{4cm}}{\raggedright 1.985e-1 (1.529e-1)} \\
		\multicolumn{1}{p{4cm}}{\raggedright Cashmere} & 
		\multicolumn{1}{p{4cm}}{\raggedright 4.118e-2 (5.539e-2) } & 
		\multicolumn{1}{p{4cm}}{\raggedright 1.018e-1*** (2.204e-2)} & 
		\multicolumn{1}{p{4cm}}{\raggedright 6.988e-2*** (1.930e-2)} \\
		\multicolumn{1}{p{4cm}}{\raggedright CHCBoys} & 
		\multicolumn{1}{p{4cm}}{\raggedright 1.356e-1 (1.413e-1)} & 
		\multicolumn{1}{p{4cm}}{\raggedright 2.277e-1*** (6.595e-2)} & 
		\multicolumn{1}{p{4cm}}{\raggedright 1.340e-1** (4.884e-2)} \\
		\multicolumn{1}{p{4cm}}{\raggedright CHCGirlsBoys} & 
		\multicolumn{1}{p{4cm}}{\raggedright 3.945e-1** (1.289e-1)} & 
		\multicolumn{1}{p{4cm}}{\raggedright 4.478e-1*** (5.480e-2)} & 
		\multicolumn{1}{p{4cm}}{\raggedright  4.010e-1*** (4.461e-2)} \\
		\multicolumn{1}{p{4cm}}{\raggedright Riccarton} & 
		\multicolumn{1}{p{4cm}}{\raggedright -7.905e-2 (9.768e-2)} & 
		\multicolumn{1}{p{4cm}}{\raggedright 2.827e-2 (4.030e-2)} & 
		\multicolumn{1}{p{4cm}}{\raggedright -6.785e-3 (3.288e-2)} \\
		\multicolumn{1}{p{4cm}}{\raggedright RiccartonCHCBoys} & 
		\multicolumn{1}{p{4cm}}{\raggedright -} & 
		\multicolumn{1}{p{4cm}}{\raggedright 3.320e-1* (1.574e-1)} & 
		\multicolumn{1}{p{4cm}}{\raggedright -} \\
		\multicolumn{1}{p{4cm}}{\raggedright RiccartonCHCGirlsBoys} & 
		\multicolumn{1}{p{4cm}}{\raggedright -} & 
		\multicolumn{1}{p{4cm}}{\raggedright -} & 
		\multicolumn{1}{p{4cm}}{\raggedright  3.681e-1* (1.527e-1)} \\
		\multicolumn{1}{p{4cm}}{\raggedright Burwood} & 
		\multicolumn{1}{p{4cm}}{\raggedright -3.579e-3 (3.608e-2)} & 
		\multicolumn{1}{p{4cm}}{\raggedright -6.371e-2*** (1.647e-2)} & 
		\multicolumn{1}{p{4cm}}{\raggedright -7.046e-2*** (1.366e-2)} \\
		\multicolumn{1}{p{4cm}}{\raggedright Fendalton} & 
		\multicolumn{1}{p{4cm}}{\raggedright 2.001e-1 (1.218e-1)} & 
		\multicolumn{1}{p{4cm}}{\raggedright 1.161e-1* (5.051e-2)} & 
		\multicolumn{1}{p{4cm}}{\raggedright 1.777e-1*** (4.121e-2)} \\ 
		\multicolumn{1}{p{4cm}}{\raggedright Linwood} & 
		\multicolumn{1}{p{4cm}}{\raggedright -1.057e-1* (5.299e-2) } & 
		\multicolumn{1}{p{4cm}}{\raggedright -1.012e-1*** (2.715e-2)} & 
		\multicolumn{1}{p{4cm}}{\raggedright -1.255e-1 *** (2.142e-2)} \\ 
		\multicolumn{1}{p{4cm}}{\raggedright Papanui} & 
		\multicolumn{1}{p{4cm}}{\raggedright 5.857e-2 (3.717e-2) } & 
		\multicolumn{1}{p{4cm}}{\raggedright 4.930e-2** (1.784e-2)} & 
		\multicolumn{1}{p{4cm}}{\raggedright 6.415e-2*** (1.498e-2)} \\ 
		\multicolumn{1}{p{4cm}}{\raggedright Riccarton} & 
		\multicolumn{1}{p{4cm}}{\raggedright 1.441e-1 (1.011e-1)} & 
		\multicolumn{1}{p{4cm}}{\raggedright 5.760e-2 (4.362e-2)} & 
		\multicolumn{1}{p{4cm}}{\raggedright 1.009e-12** (3.645e-2)} \\ 
		\multicolumn{1}{p{4cm}}{\raggedright Spreydon} & 
		\multicolumn{1}{p{4cm}}{\raggedright  2.075e-2 (4.384e-2)}& 
		\multicolumn{1}{p{4cm}}{\raggedright 1.575e-2 (2.369e-2)} & 
		\multicolumn{1}{p{4cm}}{\raggedright  8.456e-3 (1.917e-2)} \\ 
		\multicolumn{1}{p{4cm}}{\raggedright Waimairi} & 
		\multicolumn{1}{p{4cm}}{\raggedright 7.987e-2 (9.164e-2) } & 
		\multicolumn{1}{p{4cm}}{\raggedright 4.538e-2 (3.733e-2)} & 
		\multicolumn{1}{p{4cm}}{\raggedright 8.145e-2** (3.193e-2)} \\ 
		\hline\hline \\[-1.5ex]
		\multicolumn{1}{p{4cm}}{\raggedright Degrees of Freedom} & 
		\multicolumn{1}{p{4cm}}{\raggedright 316} & 
		\multicolumn{1}{p{4cm}}{\raggedright 1055} & 
		\multicolumn{1}{p{4cm}}{\raggedright 1474} \\
		\multicolumn{1}{p{4cm}}{\raggedright $R^2$} & 
		\multicolumn{1}{p{4cm}}{\raggedright 0.751} & 
		\multicolumn{1}{p{4cm}}{\raggedright 0.808} & 
		\multicolumn{1}{p{4cm}}{\raggedright 0.823} \\ 
		\multicolumn{1}{p{4cm}}{\raggedright $R^2$ adjusted} & 
		\multicolumn{1}{p{4cm}}{\raggedright 0.726} & 
		\multicolumn{1}{p{4cm}}{\raggedright 0.802} & 
		\multicolumn{1}{p{4cm}}{\raggedright 0.819} \\ 
		\multicolumn{1}{p{4cm}}{\raggedright F-statistic} & 
		\multicolumn{1}{p{4cm}}{\raggedright 30.77} & 
		\multicolumn{1}{p{4cm}}{\raggedright 126.7} & 
		\multicolumn{1}{p{4cm}}{\raggedright 195.9} \\ 
		\hline
		\caption{Table of coefficients for the model in different time periods. The legends '***' indicate significance at a 0.001 level, '**' indicates significance at a 0.01 level, '*' indicates significance at a 0.05 level, and '$\dagger$' indicates significance at a 0.1 level. }
	\end{longtable}

	Other structural attributes such as roof condition, wall condition mostly have an insignificant effect on the price of a property, except for a mixed roof condition. The reference levels used for roof condition and wall condition was 'Average'. The resulting NA coefficient values for 'WallCondition Mixed' is due to it being collinear with 'RoofCondition Mixed', while the NA coeffecient for 'WallCondition Poor' is due to the fact that there was only one observation that had that specific classification. 
	
	While roof condition being 'Mixed' seems to have a strong negative effect on a property's price, we hesitate to make any conclusions due to the limited number of observations classified in this category. There are 4, 15, and 18 observations under this classification for the 2014 to 2016, pre-lockdown, and post-lockdown periods respectively. These properties also seem to not be spatially randomly distributed.
	
	The effects of the estimated age of a property at purchase were negatively signed for all three time periods, while the squared estimated age was positively signed for all three models. This suggests that age has a depreciative effect to the price of a house up until a certain point, upon reaching a certain point the age of the property begins to have an appreciative effect, often called a vintage effect \cite{goodman1995age}. We are able to observe this phenomenon in the three models, where age reduces the price until the vintage effect comes into play once a property reaches 56, 61, and 62 years old respectively. Figure 1 illustrates this relationship clearly.
	
	\begin{figure}[htp]
		\centering
		\includegraphics[width=0.5\textwidth]{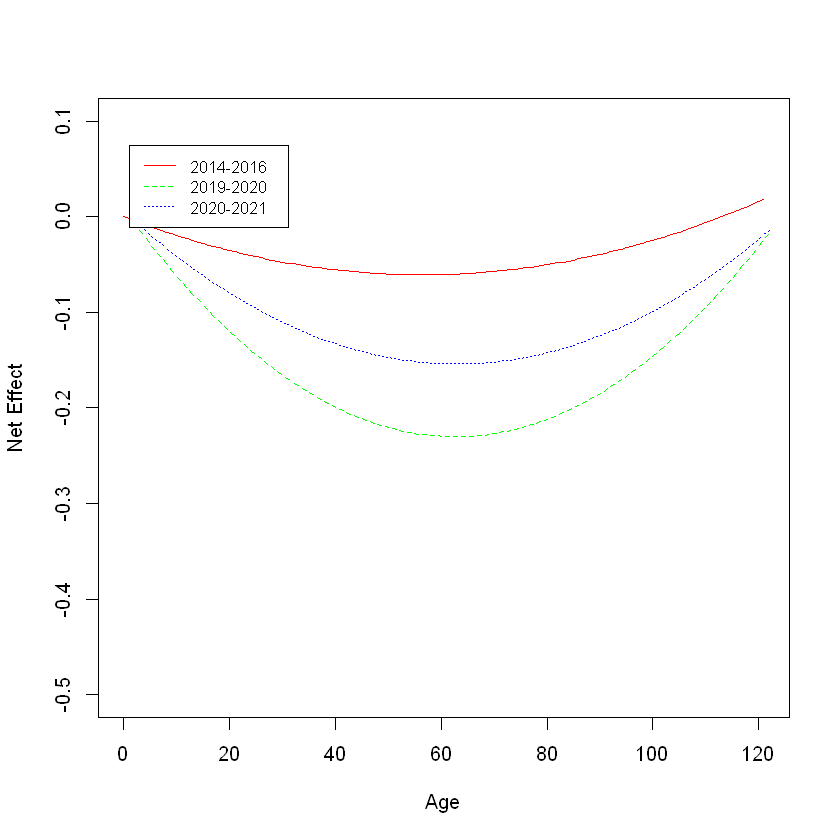}
		\caption{Net effect of age on the property price. Shows the effect of age in the depreciation of the property, and also shows the presence of the vintage effect.}
		\label{fig:age_effect}
	\end{figure}
	
	
	\subsection*{Technical Category}
	
	The effect of technical category (TC) underwent interesting changes across the three different time periods. As mentioned in Section \ref{sec:data}, properties that had classifications of TC1, TC2, and TC3 were considered. The reference level for the model is TC1, which indicates the least risk of a property experiencing damage from soil liquefaction. 
	
	In the 2014 to 2016 period, TC2 and TC3 classifications were significant at 0.01 and 0.001 levels respectively. We are able to observe that the effects are negatively signed, and TC3 has a much stronger negative effect than TC2. The effects suggest that a classification of TC3 reduces a buyer's willingness to pay by 11.9\%, and TC3 reduces it by 21.9\%. These results reflect the expected outcome as this time period is relatively close to 2011, and people were likely to exhibit caution when purchasing houses that could suffer damage in the event of another major earthquake. 
	
	For the pre-lockdown period, the effect of TC was significantly less compared to the previous time period. We found that by itself it was no longer a significant predictor of a property's price. In the case of the pre-lockdown model, only TC3 was significant at a 0.05 level. It retains its negative effect, and reduces a property's price by approximately 4.5\% which is much less compared to its effect during the 2014 to 2016 period. 
	
	In this time period, it would seem that buyers do not distinguish between properties that are classified as TC1 and TC2. The willingness to pay of buyers is also less affected if a property is classified as TC3. A possible explanation for this is as time passes people perceive the risks involved as less important since the 2011 earthquake that highlighted these risks has become quite temporaly distant. 
	
	For the post-lockdown period, the effect of TC was not what was expected. Both TC2 and TC3 were negatively signed and were significant at 0.001 and 0.05 levels respectively. However, TC2 had a stronger negative effect to a buyer's willingness to pay than if a property was classified as TC3. The effects suggest that a classification of TC2 reduces a buyer's willingness to pay by 5.3\%, and TC3 reduces it by 4\%. 
	
	However, upon changing the reference level to TC2 instead of TC1, we observed that TC3 is not considered significantly different compared to TC2. This suggests that in the post-lockdown period buyers do not distinguish between properties classified as TC2 and TC3, which is a change from the pre-lockdown period where buyers did not distinguish between TC1 and TC2.
	
	In this period the market was running hot with increasing demand outstripping supply. The change in hedonic prices for the technical categories in more likely due to a shifting market equilibrium than due to a change buyers' attitude towards  risk.

	\section{Conclusion}\label{sec:conclusion}
	
In this case study of the Christchurch housing market we applied a hedonic house price model to analyze the willingness to pay for attributes of properties in three different time periods.
	
	We observed significant shifts in hedonic prices across these three different time periods: 2014 to 2016, 2019 to 2020 pre COVID-19 lockdown, and 2020
	to 2021 post COVID-19 lockdown.
	
	In 2014 to 2016, a period relatively close to the 2011 earthquake, we found
	that a property’s locational attribute, the technical category (TC), which quantifies
	the risk of sustaining damage from soil liquefaction in the event of an earthquake,
	was a highly significant driver of house prices. The results showed that during this
	period a classification of TC3, which indicates the highest risk of damage among
	the TC classifications, reduces a buyer’s marginal willingness to pay by 21.9\% on
	average, and a classification of TC2 reduces it by 11.9\% on average. This changes
	in the following two periods which are much more temporally distant to the 2011
	earthquake. In the pre COVID-19 period,  there is no evidence that buyers distinguish between TC1 and TC2 properties, and the effect of a TC3 classification on a buyer’s marginal
	willingness to pay was significantly lower compared to the previous period. In the
	post COVID-19 lockdown period the effect of TC changed once again. There is no evidence that buyers distinguish between TC2 and TC3 properties in this period. These changes suggest
	that over the years buyers have likely started to care less about
	the risks associated with a property’s TC classification. In addition, our estimates for the effect of school zones is in agreement with the findings of \cite{mcclay2003impact}, and \cite{gibson2007house}. 
	
%
%
	
	\bibliographystyle{apalike}
	\bibliography{hedonic}

\begin{thebibliography}{}

\bibitem[Ahlfeldt and Holman, 2018]{AH:18}
Ahlfeldt, G.~M. and Holman, N. (2018).
\newblock Distinctively different: A new approach to valuing architectural
  amenities.
\newblock {\em The Economic Journal}, 128(608):1--33.

\bibitem[Atkinson et~al., 2021]{atkinson2021new}
Atkinson, J., Crampton, P., and Salmond, C. (2021).
\newblock New zealand deprivation index 2018-ta45: Upper hutt city.

\bibitem[Bishop et~al., 2020]{bishop2020best}
Bishop, K.~C., Kuminoff, N.~V., Banzhaf, H.~S., Boyle, K.~J., von Gravenitz,
  K., Pope, J.~C., Smith, V.~K., and Timmins, C.~D. (2020).
\newblock Best practices for using hedonic property value models to measure
  willingness to pay for environmental quality.
\newblock {\em Review of Environmental Economics and Policy}, 14(2):260--281.

\bibitem[Dubin, 1992]{dubin1992spatial}
Dubin, R.~A. (1992).
\newblock Spatial autocorrelation and neighborhood quality.
\newblock {\em Regional science and urban economics}, 22(3):433--452.

\bibitem[Filippova and Rehm, 2011]{filippova2011impact}
Filippova, O. and Rehm, M. (2011).
\newblock The impact of proximity to cell phone towers on residential property
  values.
\newblock {\em International Journal of Housing Markets and Analysis}.

\bibitem[Fletcher et~al., 2000]{fletcher2000heteroscedasticity}
Fletcher, M., Gallimore, P., and Mangan, J. (2000).
\newblock Heteroscedasticity in hedonic house price models.
\newblock {\em Journal of Property Research}, 17(2):93--108.

\bibitem[Gibson et~al., 2007]{gibson2007house}
Gibson, J., Sabel, C., Boe-Gibson, G., and Kim, B. (2007).
\newblock House prices and school zones: does geography matter?
\newblock In {\em Spatial Econometrics Association conference, Cambridge}.

\bibitem[Goodman and Thibodeau, 1995]{goodman1995age}
Goodman, A.~C. and Thibodeau, T.~G. (1995).
\newblock Age-related heteroskedasticity in hedonic house price equations.
\newblock {\em Journal of Housing Research}, pages 25--42.

\bibitem[Greenaway-McGrevy et~al., 2021]{greenaway2021effect}
Greenaway-McGrevy, R., Pacheco, G., and Sorensen, K. (2021).
\newblock The effect of upzoning on house prices and redevelopment premiums in
  auckland, new zealand.
\newblock {\em Urban Studies}, 58(5):959--976.

\bibitem[Helbich et~al., 2014]{helbich2014spatial}
Helbich, M., Brunauer, W., Vaz, E., and Nijkamp, P. (2014).
\newblock Spatial heterogeneity in hedonic house price models: The case of
  austria.
\newblock {\em Urban Studies}, 51(2):390--411.

\bibitem[Ikefuji et~al., 2022]{ikefuji2022earthquake}
Ikefuji, M., Laeven, R.~J., Magnus, J.~R., and Yue, Y. (2022).
\newblock Earthquake risk embedded in property prices: Evidence from five
  japanese cities.
\newblock {\em Journal of the American Statistical Association},
  117(537):82--93.

\bibitem[Lancaster, 1966]{lancaster1966new}
Lancaster, K.~J. (1966).
\newblock A new approach to consumer theory.
\newblock {\em Journal of political economy}, 74(2):132--157.

\bibitem[Lieske et~al., 2021]{lieske2021novel}
Lieske, S.~N., van~den Nouwelant, R., Han, J.~H., and Pettit, C. (2021).
\newblock A novel hedonic price modelling approach for estimating the impact of
  transportation infrastructure on property prices.
\newblock {\em Urban Studies}, 58(1):182--202.

\bibitem[Limsombunchai, 2004]{limsombunchai2004house}
Limsombunchai, V. (2004).
\newblock House price prediction: hedonic price model vs. artificial neural
  network.
\newblock In {\em New Zealand agricultural and resource economics society
  conference}, pages 25--26.

\bibitem[McClay and Harrison, 2003]{mcclay2003impact}
McClay, S. and Harrison, R. (2003).
\newblock The impact of schooling zoning on residential house prices in
  christchurch.
\newblock In {\em annual conference of the New Zealand Association of
  Economists}.

\bibitem[Palmquist, 2005]{Palmquist:05}
Palmquist, R.~B. (2005).
\newblock Chapter 16 property value models.
\newblock In Mler, K.-G. and Vincent, J.~R., editors, {\em Valuing
  Environmental Changes}, volume~2 of {\em Handbook of Environmental
  Economics}, pages 763--819. Elsevier.

\bibitem[Rehm and Filippova, 2008]{rehm2008impact}
Rehm, M. and Filippova, O. (2008).
\newblock The impact of geographically defined school zones on house prices in
  new zealand.
\newblock {\em International Journal of Housing Markets and Analysis}.

\bibitem[Rosen, 1974]{rosen1974hedonic}
Rosen, S. (1974).
\newblock Hedonic prices and implicit markets: product differentiation in pure
  competition.
\newblock {\em Journal of political economy}, 82(1):34--55.

\bibitem[Tripplett, 2004]{Triplett04}
Tripplett, J. (2004).
\newblock Handbook on hedonic indexes and quality adjustments in price indexes:
  Special application to information technology products.
\newblock {\em OECD Science, Technology and Industry Working Papers}, 2004/09.

\end{thebibliography}

	\clearpage
	\appendix
	\section{Tables}
	
	\begin{table}[!htp]
		\caption{Input data for post-lockdown period $n = 1510$}
		\centering 
		\hspace*{-2cm}\begin{tabular}{c lllll} 
			\hline\hline \\ [-1.5ex]
			\textbf{Variable} & \textbf{Data Type} & \textbf{Min} & \textbf{Max} & \textbf{Mean} & \textbf{Std. Dev.} \\\hline
			\\ [-1.5ex]
			\multicolumn{1}{p{3cm}}{\raggedright Log Real Price} & \multicolumn{1}{p{3cm}}{\raggedright Floating point} & \multicolumn{1}{p{1cm}}{\raggedright 4.709} & 
			\multicolumn{1}{p{1.5cm}}{\raggedright 7.610} & 
			\multicolumn{1}{p{1.5cm}}{\raggedright 5.294} & 
			\multicolumn{1}{p{1.5cm}}{\raggedright 0.343} \\ 
			\multicolumn{1}{p{3cm}}{\raggedright Bedrooms} & 
			\multicolumn{1}{p{3cm}}{\raggedright Integer} & 
			\multicolumn{1}{p{1.5cm}}{\raggedright 1} & 
			\multicolumn{1}{p{1.5cm}}{\raggedright 7} & 
			\multicolumn{1}{p{1.5cm}}{\raggedright -} & 
			\multicolumn{1}{p{1.5cm}}{\raggedright 0.782} \\
			\multicolumn{1}{p{3cm}}{\raggedright Bathooms} & 
			\multicolumn{1}{p{3cm}}{\raggedright Integer} & 
			\multicolumn{1}{p{1cm}}{\raggedright 1} & 
			\multicolumn{1}{p{1.5cm}}{\raggedright 3} & 
			\multicolumn{1}{p{1.5cm}}{\raggedright -} & 
			\multicolumn{1}{p{1.5cm}}{\raggedright 0.604} \\ 
			\multicolumn{1}{p{3cm}}{\raggedright Toilets} & 
			\multicolumn{1}{p{3cm}}{\raggedright Integer} & 
			\multicolumn{1}{p{1.5cm}}{\raggedright 1} & 
			\multicolumn{1}{p{1.5cm}}{\raggedright 6} & 
			\multicolumn{1}{p{1.5cm}}{\raggedright -} & 
			\multicolumn{1}{p{1.5cm}}{\raggedright 0.802} \\ 
			\multicolumn{1}{p{3cm}}{\raggedright Carparks} & 
			\multicolumn{1}{p{3cm}}{\raggedright Integer} & 
			\multicolumn{1}{p{1.5cm}}{\raggedright 1} & 
			\multicolumn{1}{p{1.5cm}}{\raggedright 8} & 
			\multicolumn{1}{p{1.5cm}}{\raggedright -} & 
			\multicolumn{1}{p{1.5cm}}{\raggedright 1.870} \\
			\multicolumn{1}{p{3cm}}{\raggedright Floor Area} & 
			\multicolumn{1}{p{3cm}}{\raggedright Floating Point} & 
			\multicolumn{1}{p{1.5cm}}{\raggedright 68.0$m^2$} & 
			\multicolumn{1}{p{1.5cm}}{\raggedright 563.0$m^2$} & 
			\multicolumn{1}{p{1.5cm}}{\raggedright 163.015$m^2$} & 
			\multicolumn{1}{p{1cm}}{\raggedright 66.0} \\ 
			\multicolumn{1}{p{3cm}}{\raggedright Land Area} & 
			\multicolumn{1}{p{3cm}}{\raggedright Floating Point} & 
			\multicolumn{1}{p{1.5cm}}{\raggedright 217.0$m^2$} & 
			\multicolumn{1}{p{1.5cm}}{\raggedright 2611.0$m^2$} & 
			\multicolumn{1}{p{1.5cm}}{\raggedright 693.9$m^2$} & 
			\multicolumn{1}{p{1.56cm}}{\raggedright 211.386} \\ 
			\multicolumn{1}{p{3cm}}{\raggedright Roof Condition} & 
			\multicolumn{1}{p{3cm}}{\raggedright Categorical} & 
			\multicolumn{1}{p{1.5cm}}{\raggedright -} & 
			\multicolumn{1}{p{1.5cm}}{\raggedright -} & 
			\multicolumn{1}{p{1.5cm}}{\raggedright -} & 
			\multicolumn{1}{p{1.56cm}}{\raggedright -} \\ 
			\multicolumn{1}{p{3cm}}{\raggedright View from Property} & 
			\multicolumn{1}{p{3cm}}{\raggedright Categorical} & 
			\multicolumn{1}{p{1.5cm}}{\raggedright -} & 
			\multicolumn{1}{p{1.5cm}}{\raggedright -} & 
			\multicolumn{1}{p{1.5cm}}{\raggedright -} & 
			\multicolumn{1}{p{1.56cm}}{\raggedright -} \\
			\multicolumn{1}{p{3cm}}{\raggedright Wall Condition} & 
			\multicolumn{1}{p{3cm}}{\raggedright Categorical} & 
			\multicolumn{1}{p{1.5cm}}{\raggedright -} & 
			\multicolumn{1}{p{1.5cm}}{\raggedright -} & 
			\multicolumn{1}{p{1.5cm}}{\raggedright -} & 
			\multicolumn{1}{p{1.56cm}}{\raggedright -} \\ 
			\multicolumn{1}{p{3cm}}{\raggedright Dist. CBD} & 
			\multicolumn{1}{p{3cm}}{\raggedright Floating point} & 
			\multicolumn{1}{p{1.5cm}}{\raggedright 1.579$km$} & 
			\multicolumn{1}{p{1.5cm}}{\raggedright 14.301$km$} & 
			\multicolumn{1}{p{1.5cm}}{\raggedright 7.3$km$} & 
			\multicolumn{1}{p{1.56cm}}{\raggedright 1.990} \\ 
			\multicolumn{1}{p{3cm}}{\raggedright Dist. to Secondary} & 
			\multicolumn{1}{p{3cm}}{\raggedright Floating point} & 
			\multicolumn{1}{p{1.5cm}}{\raggedright 0.085$km$} & 
			\multicolumn{1}{p{1.5cm}}{\raggedright 9.874$km$} & 
			\multicolumn{1}{p{1.5cm}}{\raggedright 1.741$km$} & 
			\multicolumn{1}{p{1.56cm}}{\raggedright 1.209} \\
			\multicolumn{1}{p{3cm}}{\raggedright Technical Category} & 
			\multicolumn{1}{p{3cm}}{\raggedright Categorical} & 
			\multicolumn{1}{p{1.5cm}}{\raggedright -} & 
			\multicolumn{1}{p{1.5cm}}{\raggedright -} & 
			\multicolumn{1}{p{1.5cm}}{\raggedright -} & 
			\multicolumn{1}{p{1.56cm}}{\raggedright -} \\  
			\multicolumn{1}{p{3cm}}{\raggedright Age Purchased} & 
			\multicolumn{1}{p{3cm}}{\raggedright Integer} & 
			\multicolumn{1}{p{1.5cm}}{\raggedright 0} & 
			\multicolumn{1}{p{1.5cm}}{\raggedright 121} & 
			\multicolumn{1}{p{1.5cm}}{\raggedright 50.39} & 
			\multicolumn{1}{p{1.56cm}}{\raggedright 25.711} \\ 
			\multicolumn{1}{p{3cm}}{\raggedright NZ Deprivation Score} & 
			\multicolumn{1}{p{3cm}}{\raggedright Integer} & 
			\multicolumn{1}{p{1.5cm}}{\raggedright 839} & 
			\multicolumn{1}{p{1.5cm}}{\raggedright 1365} & 
			\multicolumn{1}{p{1.5cm}}{\raggedright 969.4} & 
			\multicolumn{1}{p{1.56cm}}{\raggedright 70.113} \\  
			\multicolumn{1}{p{3cm}}{\raggedright School Zone} & 
			\multicolumn{1}{p{3cm}}{\raggedright Categorical} & 
			\multicolumn{1}{p{1.5cm}}{\raggedright -} & 
			\multicolumn{1}{p{1.5cm}}{\raggedright -} & 
			\multicolumn{1}{p{1.5cm}}{\raggedright -} & 
			\multicolumn{1}{p{1.56cm}}{\raggedright -} \\ 
			\multicolumn{1}{p{3cm}}{\raggedright Ward} & 
			\multicolumn{1}{p{1.5cm}}{\raggedright Categorical} & 
			\multicolumn{1}{p{1.5cm}}{\raggedright -} & 
			\multicolumn{1}{p{1.5cm}}{\raggedright -} & 
			\multicolumn{1}{p{1.56cm}}{\raggedright -} \\ 
			\multicolumn{1}{p{4cm}}{\raggedright (Age Purchased)$^2$} & 
			\multicolumn{1}{p{1.5cm}}{\raggedright Integer} & 
			\multicolumn{1}{p{1.5cm}}{\raggedright 0} & 
			\multicolumn{1}{p{1.5cm}}{\raggedright 14641} & 
			\multicolumn{1}{p{1.5cm}}{\raggedright 3199.708} &
			\multicolumn{1}{p{1.5cm}}{\raggedright 2790.211} \\ 
			\hline
		\end{tabular}\hspace*{-2cm}
		\label{tab:input_data_2014}
	\end{table}
	
	\clearpage
	
	\begin{table}[!htp]
		\caption{Input data for 2014-2015 $n=349$}
		\centering 
		\hspace*{-2cm}\begin{tabular}{c lllll} 
			\hline\hline \\ [-1.5ex]
			\textbf{Variable} & \textbf{Data Type} & \textbf{Min} & \textbf{Max} & \textbf{Mean} & \textbf{Std. Dev.} \\\hline
			\\ [-1.5ex]
			\multicolumn{1}{p{3cm}}{\raggedright Log Real Price} & 
			\multicolumn{1}{p{3cm}}{\raggedright Floating point} & 
			\multicolumn{1}{p{1cm}}{\raggedright 4.728} & 
			\multicolumn{1}{p{1.5cm}}{\raggedright 6.794} & 
			\multicolumn{1}{p{1.5cm}}{\raggedright 5.497} & 
			\multicolumn{1}{p{1.5cm}}{\raggedright 0.346} \\ 
			\multicolumn{1}{p{3cm}}{\raggedright Bedrooms} & 
			\multicolumn{1}{p{3cm}}{\raggedright Integer} & 
			\multicolumn{1}{p{1.5cm}}{\raggedright 2} & 
			\multicolumn{1}{p{1.5cm}}{\raggedright 6} & 
			\multicolumn{1}{p{1.5cm}}{\raggedright -} & 
			\multicolumn{1}{p{1.5cm}}{\raggedright 0.751} \\
			\multicolumn{1}{p{3cm}}{\raggedright Bathooms} & 
			\multicolumn{1}{p{3cm}}{\raggedright Integer} & 
			\multicolumn{1}{p{1cm}}{\raggedright 1} & 
			\multicolumn{1}{p{1.5cm}}{\raggedright 3} & 
			\multicolumn{1}{p{1.5cm}}{\raggedright -} & 
			\multicolumn{1}{p{1.5cm}}{\raggedright 0.628} \\ 
			\multicolumn{1}{p{3cm}}{\raggedright Toilets} & 
			\multicolumn{1}{p{3cm}}{\raggedright Integer} & 
			\multicolumn{1}{p{1.5cm}}{\raggedright 1} & 
			\multicolumn{1}{p{1.5cm}}{\raggedright 4} & 
			\multicolumn{1}{p{1.5cm}}{\raggedright -} & 
			\multicolumn{1}{p{1.5cm}}{\raggedright 0.793} \\ 
			\multicolumn{1}{p{3cm}}{\raggedright Carparks} & 
			\multicolumn{1}{p{3cm}}{\raggedright Integer} & 
			\multicolumn{1}{p{1.5cm}}{\raggedright 1} & 
			\multicolumn{1}{p{1.5cm}}{\raggedright 5} & 
			\multicolumn{1}{p{1.5cm}}{\raggedright -} & 
			\multicolumn{1}{p{1.5cm}}{\raggedright 0.604} \\
			\multicolumn{1}{p{3cm}}{\raggedright Floor Area} & 
			\multicolumn{1}{p{3cm}}{\raggedright Floating Point} & 
			\multicolumn{1}{p{1.5cm}}{\raggedright 80.0$m^2$} & 
			\multicolumn{1}{p{1.5cm}}{\raggedright 430.0$m^2$} & 
			\multicolumn{1}{p{1.5cm}}{\raggedright 153.1$m^2$} & 
			\multicolumn{1}{p{1cm}}{\raggedright 62.199} \\ 
			\multicolumn{1}{p{3cm}}{\raggedright Land Area} & 
			\multicolumn{1}{p{3cm}}{\raggedright Floating Point} & 
			\multicolumn{1}{p{1.5cm}}{\raggedright 284.0$m^2$} & 
			\multicolumn{1}{p{1.5cm}}{\raggedright 2529.0$m^2$} & 
			\multicolumn{1}{p{1.5cm}}{\raggedright 686.5$m^2$} & 
			\multicolumn{1}{p{1.56cm}}{\raggedright 176.013} \\ 
			\multicolumn{1}{p{3cm}}{\raggedright Position of Property} & 
			\multicolumn{1}{p{3cm}}{\raggedright Categorical} & 
			\multicolumn{1}{p{1.5cm}}{\raggedright -} & 
			\multicolumn{1}{p{1.5cm}}{\raggedright -} & 
			\multicolumn{1}{p{1.5cm}}{\raggedright -} & 
			\multicolumn{1}{p{1.56cm}}{\raggedright -} \\ 
			\multicolumn{1}{p{3cm}}{\raggedright Roof Condition} & 
			\multicolumn{1}{p{3cm}}{\raggedright Categorical} & 
			\multicolumn{1}{p{1.5cm}}{\raggedright -} & 
			\multicolumn{1}{p{1.5cm}}{\raggedright -} & 
			\multicolumn{1}{p{1.5cm}}{\raggedright -} & 
			\multicolumn{1}{p{1.56cm}}{\raggedright -} \\ 
			\multicolumn{1}{p{3cm}}{\raggedright View from Property} & 
			\multicolumn{1}{p{3cm}}{\raggedright Categorical} & 
			\multicolumn{1}{p{1.5cm}}{\raggedright -} & 
			\multicolumn{1}{p{1.5cm}}{\raggedright -} & 
			\multicolumn{1}{p{1.5cm}}{\raggedright -} & 
			\multicolumn{1}{p{1.56cm}}{\raggedright -} \\
			\multicolumn{1}{p{3cm}}{\raggedright Wall Condition} & 
			\multicolumn{1}{p{3cm}}{\raggedright Categorical} & 
			\multicolumn{1}{p{1.5cm}}{\raggedright -} & 
			\multicolumn{1}{p{1.5cm}}{\raggedright -} & 
			\multicolumn{1}{p{1.5cm}}{\raggedright -} & 
			\multicolumn{1}{p{1.56cm}}{\raggedright -} \\ 
			\multicolumn{1}{p{3cm}}{\raggedright Dist. CBD} & 
			\multicolumn{1}{p{3cm}}{\raggedright Floating point} & 
			\multicolumn{1}{p{1.5cm}}{\raggedright 1.888$km$} & 
			\multicolumn{1}{p{1.5cm}}{\raggedright 11.766$km$} & 
			\multicolumn{1}{p{1.5cm}}{\raggedright 5.860$km$} & 
			\multicolumn{1}{p{1.56cm}}{\raggedright 1.914} \\ 
			\multicolumn{1}{p{3cm}}{\raggedright Dist. to Secondary} & 
			\multicolumn{1}{p{3cm}}{\raggedright Floating point} & 
			\multicolumn{1}{p{1.5cm}}{\raggedright 0.097$km$} & 
			\multicolumn{1}{p{1.5cm}}{\raggedright 8.762$km$} & 
			\multicolumn{1}{p{1.5cm}}{\raggedright 1.665$km$} & 
			\multicolumn{1}{p{1.56cm}}{\raggedright 1.272} \\
			\multicolumn{1}{p{3cm}}{\raggedright Technical Category} & 
			\multicolumn{1}{p{3cm}}{\raggedright Categorical} & 
			\multicolumn{1}{p{1.5cm}}{\raggedright -} & 
			\multicolumn{1}{p{1.5cm}}{\raggedright -} & 
			\multicolumn{1}{p{1.5cm}}{\raggedright -} & 
			\multicolumn{1}{p{1.56cm}}{\raggedright -} \\  
			\multicolumn{1}{p{3cm}}{\raggedright Age Purchased} & 
			\multicolumn{1}{p{3cm}}{\raggedright Integer} & 
			\multicolumn{1}{p{1.5cm}}{\raggedright 2} & 
			\multicolumn{1}{p{1.5cm}}{\raggedright 120} & 
			\multicolumn{1}{p{1.5cm}}{\raggedright 52.47} & 
			\multicolumn{1}{p{1.56cm}}{\raggedright 24.103} \\ 
			\multicolumn{1}{p{3cm}}{\raggedright NZ Deprivation Score (2013)} & 
			\multicolumn{1}{p{3cm}}{\raggedright Integer} & 
			\multicolumn{1}{p{1.5cm}}{\raggedright 855} & 
			\multicolumn{1}{p{1.5cm}}{\raggedright 1286} & 
			\multicolumn{1}{p{1.5cm}}{\raggedright 958.1} & 
			\multicolumn{1}{p{1.56cm}}{\raggedright 70.170} \\  
			\multicolumn{1}{p{3cm}}{\raggedright School Zone} & 
			\multicolumn{1}{p{3cm}}{\raggedright Categorical} & 
			\multicolumn{1}{p{1.5cm}}{\raggedright -} & 
			\multicolumn{1}{p{1.5cm}}{\raggedright -} & 
			\multicolumn{1}{p{1.5cm}}{\raggedright -} & 
			\multicolumn{1}{p{1.56cm}}{\raggedright -} \\ 
			\multicolumn{1}{p{3cm}}{\raggedright Ward} & 
			\multicolumn{1}{p{1.5cm}}{\raggedright Categorical} & 
			\multicolumn{1}{p{1.5cm}}{\raggedright -} & 
			\multicolumn{1}{p{1.5cm}}{\raggedright -} & 
			\multicolumn{1}{p{1.56cm}}{\raggedright -} \\ 
			\multicolumn{1}{p{4cm}}{\raggedright (Age Purchased)$^2$} & 
			\multicolumn{1}{p{1.5cm}}{\raggedright Integer} & 
			\multicolumn{1}{p{1.5cm}}{\raggedright 0} & 
			\multicolumn{1}{p{1.5cm}}{\raggedright 14161} & 
			\multicolumn{1}{p{1.5cm}}{\raggedright 332.372} &
			\multicolumn{1}{p{1.5cm}}{\raggedright 2641.848} \\ 
			\hline
		\end{tabular}\hspace*{-2cm}
		\label{tab:input_data_post}
	\end{table}
	
	\clearpage
	
	\section{Discussion of Locational Attributes} \label{app:discussion}

	Location related features such as distance to the central business district (CBD), and the distance to a secondary school have consistent signs across the three different time periods and are significant to a 0.001 level. The negative effect of a property's distance to the CBD is expected as property prices tend to increase based on how much more access to commercial amenities and work areas it has. We observe sizeable changes in the effect between time periods. The buyers' willingness to pay for a property decreases by approximately 4.2\%, 1.6\%, and 2.9\% for every additional kilometer farther away it is from the CBD for each time period respectively. 
	
	The positive effect of distance to a secondary school at first glance is unintuitive, as school accessibility in terms of distance is expected to have a positive influence on a property's price. However, if we take into account the context of Christchurch, or even New Zealand as a whole, being a place where privately driven vehicles are the most popular form of commuting, the effect becomes understandable. Distance to a school matters less as students simply are driven by their parents to school. What we could be observing are the negative effects of being close to a secondary to school, such as increased traffic activity in the area at certain hours of the day due to students coming and going. This can also have measurable impacts on other aspects of quality of life, such as noise levels, overall air quality, etc.
	
	The New Zealand deprivation index was significant at a 0.001 level for all three models, and maintained a consistent negative effect throughout. It is important to note that effect size for the deprivation index in the model for the  2014 to 2016 period is not directly comparable with that of the other two models since it uses the New Zealand deprivation index for 2013 instead of the one for 2018. There was no notable difference in the effect of the deprivation index between the two models for the pre-lockdown and post-lockdown periods. The effect size suggests that, all else equal, a property located in an area that has a deprivation score one unit higher would be 0.064\% lower in price.
	
	School zones in New Zealand are strong drivers of house price, as shown in \cite{mcclay2003impact} ,\cite{gibson2007house}, and \cite{rehm2008impact}. We observe similar findings here, further verifying the results in \cite{gibson2007house}. 
	
	The reference level for the school zone variable is 'Other', which indicates that a property does not belong in any of the five main school zones considered i.e. Burnside Highschool, Christchurch Boys' Highschool, Christchurch Girls' Highschool, Riccaroton Highschool, and Cashmere Highschool. 
	
	There is some difficulty in comparing the effects of school zones across different models, as it is entirely possible for a property to belong to multiple zones. This also leads to the possibility that, for a given time period, no property sold falls in a particular classification. This leads to the lack of coefficients for particular combinations of school zones. One thing to note is that the Christchurch Girls' Highschool zone is completely enclosed in the Christchurch Boys' Highschool zone, which effectively means that there are no properties that solely belong to the Christchurch Girls' school zone. In this case, 'CHCGirlsBoys' effectively stands as the Christchurch Girls' school zone. 
	
	For the period 2014 to 2016, 'CHCGirlsBoys' was significantly different from the reference category at a 0.01 level. It was the only zone that was significantly different from the reference level for this period. The effect suggests that, all else held equal, a buyer is willing to pay up to a 48.3\% premium for property that is in the Christchurch Girls' Highschool zone as compared to an identically featured house that is not in any of the five considered school zones. This translates to approximately a \$239,000 premium in price based on the median prices of properties that do not belong to any of the five school zones. 
	
	The lack of other significantly different school zones in the period 2014 to 2016 departs from the results seen in \cite{mcclay2003impact}, and \cite{gibson2007house}, where Burnside Highschool, Christchurch Boys' Highschool, and Riccarton Highschool zones all have significantly different effects on the price compared to the reference level. The change we observe compared to the earlier studies may be due to some lingering effects of the 2011 Christchurch earthquake, where home buyers prioritize different characteristics such as the Technical Category classification of the property.
	
	For the pre-lockdown period, all but the Riccarton Highschool zone had significantly different effects on the price compared to the reference level. The school zone with the highest premium once again is that of Christchurch Girls' Highschool, where consumers are willing to pay up to 56.5\% more for a similarly featured property that is not within any of the five school zones. The other premiums that buyers are willing to pay are 11.5\%, 10.7\%, and 25.6\% for properties in Burnside Highschool, Cashmere Highschool, and Christchurch Boys' Highschool zones only, respectively.
	
	During the post-lockdown period, less school zones, or combinations thereof, had significantly different effects compared to the reference level. Once again, Christchurch Girls' Highschool commands the highest premium where buyers are willing to pay 49.3\% more compared to a non-zoned, identically featured house. The pre-lockdown estimates agree with the results of \cite{mcclay2003impact}, and \cite{gibson2007house} in terms of the significance of the effect of different school zones, while post-lockdown results do not. It could be the case that because the market has been much hotter in the months after the lockdown, the implicit price of desirable school zones in the new market equilibrium is decreasing, i.e. school zones play a smaller role in the determination of a buyers' decision to purchase a property relative to other factors. 
	
	The results for the effect sizes for which ward a property belongs in were generally expected. The reference level was 'Other', which indicates properties that do not belong in any of the seven considered wards. For the period 2014 to 2016, Linwood's effect was negatively signed and significantly different from the reference category's effect on price at a 0.05 level. It was the only ward that had a significantly different effect compared to the reference level. The effect suggests that the price for a property located in Linwood decreases by approximately 10\% compared to a similarly featured property that is outside the seven wards. 
	
	The results for the pre-lockdown period suggest that the effect of which ward a property is located in assumed a greater role when it comes to determining a buyer's willingness to pay compared to the 2014 to 2016 period. 
	
	During this period there were four wards that had significant effects on the price, namely Burwood, Linwood, Fendalton and Papanui. Linwood maintains a strong negative effect at a 0.001 level of significance which suggests that a property being in Linwood decreases the price by approximately 9.6\%. A property being located in Fendalton had a strong positive effect which was significant to a 0.05 level. The effect suggests that the price for a property in Fendalton is 12.3\% higher compared to one that is not within the seven wards. Being located in Burwood decreases the price of a property by approximately 6.1\%, and being located in Papanui increases the price of a property approximately 5\% compared to the reference category.
	
	We observe a significant shift in the importance of ward classification in the post-lockdown period compared to the previous two periods. Every ward classification except for Spreydon had a significant effect on a buyer's willingness to pay for a property. This shift in preferences aligns with the stipulation made above, where buyers, post-lockdown, experienced a change in priorities in terms of the qualities sought after in a property as well as shifts in implicit prices due to increased demand. We suspect that there are two main explanations for this shift in preferences. First, as mentioned in Section \ref{sec:data}, new school zone demarcations were intially made known in 2019 which were set to take effect some time in 2022. What we might be observing is the effect of these boundary changes, as it entails significant changes in the scope of particular school zones compared to the data used for this paper, which would be the zoning schedule established in 2010. For example, Burnside Highschool's zone effectively doubles when these changes take effect, which could possibly explain the insignificance of the effect of Burnside Highschool in the post-lockdown period.
	
	Another plausible explanation is that given the market is hot the amount of transactions conducted by investors, not would-be homeowners, is likely to be higher than it would be if it were a buyer's market. 
	
	In the post-lockdown period, the Fendalton ward once again commands the highest premium. Its effect is significant at a 0.001 level and suggests that buyers are willing to pay up to 19.4\% more if a property is located in the Fendalton ward compared to a property that is not located in any of the other six wards. Linwood and Burwood retain their strong negative effects which are significant at a 0.001 level. Their effects suggests that prices of properties located in Linwood and Burwood decrease by approximately 11.8\%, and 6.8\% respectively.

	\section{A time dummy model}\label{app:timedummy}
	
	For the results presented in this paper we used real prices. The prices where deflated with the house price index provided by the Reserve Bank of New Zealand. An alternative approach is a model using nominal prices and time dummies to account for systematic trends in prices. In fact, models of this type are often used to generate a price index in the first place as described in \cite{Triplett04}. Table \ref{tab:dummies} presents results for this approach in our context with quarterly dummies. The results generally agree with the model without dummies using real prices. 
	\setlength\LTleft{0cm}
	\setlength\LTright{0pt plus 1fill minus 1fill}%
	\small
	\setlength\LTleft{-2cm}
	\setlength\LTright{0pt plus 1fill minus 1fill}
	\begin{longtable}{c lll} 
		\caption{Hedonic regression estimates by time period with quarterly time dummies}
		\label{tab:dummies} \\
		\hline\hline \\ [-1.5ex]
		\textbf{Variable} & \textbf{2014-2016} & \textbf{2019-2020} & \textbf{2020-2021} \\
		& Coefficient (SE) & Coefficient (SE) &  Coefficient (SE) \\
		\hline
		\\ [-1.5ex]
		\multicolumn{1}{p{4cm}}{\raggedright Intercept} & 
		\multicolumn{1}{p{4cm}}{\raggedright 1.381*** (2.241e-1)} & 
		\multicolumn{1}{p{4cm}}{\raggedright 1.347*** (1.664e-1)} & 
		\multicolumn{1}{p{4cm}}{\raggedright 1.361*** (8.197e-2 )} \\
		\multicolumn{1}{p{4cm}}{\raggedright Quarter 2} & 
		\multicolumn{1}{p{4cm}}{\raggedright -} & 
		\multicolumn{1}{p{4cm}}{\raggedright -3.952e-2 (2.755e-2)} & 
		\multicolumn{1}{p{4cm}}{\raggedright -1.201e-1*** (1.148e-2)} \\
		\multicolumn{1}{p{4cm}}{\raggedright Quarter 3} & 
		\multicolumn{1}{p{4cm}}{\raggedright -} & 
		\multicolumn{1}{p{4cm}}{\raggedright -1.434e-2 (1.546e-2)} & 
		\multicolumn{1}{p{4cm}}{\raggedright -7.595e-2*** (1.099e-2)} \\
		\multicolumn{1}{p{4cm}}{\raggedright Quarter 4} & 
		\multicolumn{1}{p{4cm}}{\raggedright -} & 
		\multicolumn{1}{p{4cm}}{\raggedright -9.093e-4 (1.094e-2)} & 
		\multicolumn{1}{p{4cm}}{\raggedright -1.468e-3 (1.047e-2)} \\ 
		\multicolumn{1}{p{4cm}}{\raggedright Quarter 1-2015} & 
		\multicolumn{1}{p{4cm}}{\raggedright -5.323e-3 (4.677e-2)} & 
		\multicolumn{1}{p{4cm}}{\raggedright -} & 
		\multicolumn{1}{p{4cm}}{\raggedright -} \\ 
		\multicolumn{1}{p{4cm}}{\raggedright Quarter 2-2014} & 
		\multicolumn{1}{p{4cm}}{\raggedright -5.595e-2  (4.709e-2)} & 
		\multicolumn{1}{p{4cm}}{\raggedright -} & 
		\multicolumn{1}{p{4cm}}{\raggedright -} \\ 
		\multicolumn{1}{p{4cm}}{\raggedright Quarter 2-2015} & 
		\multicolumn{1}{p{4cm}}{\raggedright -3.348e-2 (4.420e-2)} & 
		\multicolumn{1}{p{4cm}}{\raggedright -} & 
		\multicolumn{1}{p{4cm}}{\raggedright -} \\ 
		\multicolumn{1}{p{4cm}}{\raggedright Quarter 3-2014} & 
		\multicolumn{1}{p{4cm}}{\raggedright -5.264e-3 (4.483e-2)} & 
		\multicolumn{1}{p{4cm}}{\raggedright -} & 
		\multicolumn{1}{p{4cm}}{\raggedright -} \\ 
		\multicolumn{1}{p{4cm}}{\raggedright Quarter 3-2015} & 
		\multicolumn{1}{p{4cm}}{\raggedright-2.400e-2 (4.525e-2)} & 
		\multicolumn{1}{p{4cm}}{\raggedright -} & 
		\multicolumn{1}{p{4cm}}{\raggedright -} \\ 
		\multicolumn{1}{p{4cm}}{\raggedright Quarter 4-2014} & 
		\multicolumn{1}{p{4cm}}{\raggedright -1.940e-2 (4.343e-2)} & 
		\multicolumn{1}{p{4cm}}{\raggedright -} & 
		\multicolumn{1}{p{4cm}}{\raggedright -} \\ 
		\multicolumn{1}{p{4cm}}{\raggedright Quarter 4-2015} & 
		\multicolumn{1}{p{4cm}}{\raggedright 3.946e-2 (4.616e-2)} & 
		\multicolumn{1}{p{4cm}}{\raggedright -} & 
		\multicolumn{1}{p{4cm}}{\raggedright -} \\ 
		\multicolumn{1}{p{4cm}}{\raggedright Bedrooms} & 
		\multicolumn{1}{p{4cm}}{\raggedright 1.319e-2 (3.197e-2)} & 
		\multicolumn{1}{p{4cm}}{\raggedright -5.517e-3 (1.135e-2 ) } & 
		\multicolumn{1}{p{4cm}}{\raggedright -5.443e-2*** (1.054e-2)} \\
		\multicolumn{1}{p{4cm}}{\raggedright Bathrooms} & 
		\multicolumn{1}{p{4cm}}{\raggedright -2.323e-2  (3.943e-2)} & 
		\multicolumn{1}{p{4cm}}{\raggedright 5.672e-3 (1.608e-2)} & 
		\multicolumn{1}{p{4cm}}{\raggedright 2.139e-2 $\dagger$ (1.290e-2)} \\ 
		\multicolumn{1}{p{4cm}}{\raggedright Toilets} & 
		\multicolumn{1}{p{4cm}}{\raggedright 5.570e-3 (3.605e-2)} & 
		\multicolumn{1}{p{4cm}}{\raggedright 2.252e-2  (1.338e-2)} & 
		\multicolumn{1}{p{4cm}}{\raggedright 3.162e-2** (1.081e-2)} \\ 
		\multicolumn{1}{p{4cm}}{\raggedright Carparks} & 
		\multicolumn{1}{p{4cm}}{\raggedright 3.271e-3  (2.089e-2)} & 
		\multicolumn{1}{p{4cm}}{\raggedright 3.585e-3 (9.361e-3)} & 
		\multicolumn{1}{p{4cm}}{\raggedright 4.090e-3 (7.016e-3)} \\
		\multicolumn{1}{p{4cm}}{\raggedright Floor Area} & 
		\multicolumn{1}{p{4cm}}{\raggedright 2.943e-3***(3.950e-4) } & 
		\multicolumn{1}{p{4cm}}{\raggedright 2.203e-3*** (1.572e-4)} & 
		\multicolumn{1}{p{4cm}}{\raggedright 2.370e-3*** (1.276e-4)} \\ 
		\multicolumn{1}{p{4cm}}{\raggedright Land Area} & 
		\multicolumn{1}{p{4cm}}{\raggedright 2.491e-4 (7.209e-5)} & 
		\multicolumn{1}{p{4cm}}{\raggedright 2.103e-4***(2.590e-5)} & 
		\multicolumn{1}{p{4cm}}{\raggedright 2.633e-4*** (2.180e-5)} \\ 
		\multicolumn{1}{p{4cm}}{\raggedright RoofCondition Fair} & 
		\multicolumn{1}{p{4cm}}{\raggedright 2.079e-1 (2.475e-1)} & 
		\multicolumn{1}{p{4cm}}{\raggedright 2.958e-2 (7.482e-2 ) } & 
		\multicolumn{1}{p{4cm}}{\raggedright -3.199e-2 (6.465e-2) } \\
		\multicolumn{1}{p{4cm}}{\raggedright RoofCondition Good} & 
		\multicolumn{1}{p{4cm}}{\raggedright -1.970e-2  (5.770e-2) } & 
		\multicolumn{1}{p{4cm}}{\raggedright -2.847e-3  (2.364e-2) } & 
		\multicolumn{1}{p{4cm}}{\raggedright 1.134e-2 (1.896e-2)} \\ 
		\multicolumn{1}{p{4cm}}{\raggedright RoofCondition Mixed} & 
		\multicolumn{1}{p{4cm}}{\raggedright -5.561e-1*** (1.113e-1)} & 
		\multicolumn{1}{p{4cm}}{\raggedright -1.127e-1** (3.809e-2) } & 
		\multicolumn{1}{p{4cm}}{\raggedright -1.824e-1*** (4.058e-2)} \\
		\multicolumn{1}{p{4cm}}{\raggedright RoofCondition Poor} & 
		\multicolumn{1}{p{4cm}}{\raggedright -} & 
		\multicolumn{1}{p{4cm}}{\raggedright-1.936e-1 (1.551e-1)} & 
		\multicolumn{1}{p{4cm}}{\raggedright 9.099e-3 (1.505e-1)} \\ 
		\multicolumn{1}{p{4cm}}{\raggedright View - Has view } & 
		\multicolumn{1}{p{4cm}}{\raggedright 3.914e-2 (7.063e-2) } & 
		\multicolumn{1}{p{4cm}}{\raggedright 2.002e-2  (2.567e-2)} & 
		\multicolumn{1}{p{4cm}}{\raggedright 4.182e-2$\dagger$ (2.280e-2)} \\
		\multicolumn{1}{p{4cm}}{\raggedright WallCondition Fair} & 
		\multicolumn{1}{p{4cm}}{\raggedright -2.015e-1  (2.045e-1) } & 
		\multicolumn{1}{p{4cm}}{\raggedright -8.691e-2 (6.663e-2)} & 
		\multicolumn{1}{p{4cm}}{\raggedright -3.115e-2 (5.202e-2) } \\
		\multicolumn{1}{p{4cm}}{\raggedright WallCondition Good} & 
		\multicolumn{1}{p{4cm}}{\raggedright 4.621e-2 (5.669e-2)} & 
		\multicolumn{1}{p{4cm}}{\raggedright 7.452e-3  (2.365e-2)} & 
		\multicolumn{1}{p{4cm}}{\raggedright 2.401e-2 (1.864e-2)} \\ 
		\multicolumn{1}{p{4cm}}{\raggedright WallCondition Mixed} & 
		\multicolumn{1}{p{4cm}}{\raggedright NA} & 
		\multicolumn{1}{p{4cm}}{\raggedright NA} & 
		\multicolumn{1}{p{4cm}}{\raggedright NA} \\ 
		\multicolumn{1}{p{4cm}}{\raggedright WallCondition Poor} & 
		\multicolumn{1}{p{4cm}}{\raggedright -} & 
		\multicolumn{1}{p{4cm}}{\raggedright -8.832e-2 (1.547e-1) } & 
		\multicolumn{1}{p{4cm}}{\raggedright NA} \\ 
		\multicolumn{1}{p{4cm}}{\raggedright Dist. CBD} & 
		\multicolumn{1}{p{4cm}}{\raggedright -4.389e-2*** (9.276e-3)} & 
		\multicolumn{1}{p{4cm}}{\raggedright -1.632e-2*** (4.454e-3)} & 
		\multicolumn{1}{p{4cm}}{\raggedright -2.713e-2*** (3.711e-3)} \\ 
		\multicolumn{1}{p{4cm}}{\raggedright Dist. to Secondary} & 
		\multicolumn{1}{p{4cm}}{\raggedright 3.514e-2** (1.298e-2)} & 
		\multicolumn{1}{p{4cm}}{\raggedright 2.858e-2*** (5.767e-3)} & 
		\multicolumn{1}{p{4cm}}{\raggedright 4.112e-2*** (4.949e-3)} \\  
		\multicolumn{1}{p{4cm}}{\raggedright TC2} & 
		\multicolumn{1}{p{4cm}}{\raggedright -4.404e-2 (3.399e-2)} & 
		\multicolumn{1}{p{4cm}}{\raggedright -3.071e-2 (1.894e-2 )} & 
		\multicolumn{1}{p{4cm}}{\raggedright -4.129e-2** (1.567e-2 )} \\ 
		\multicolumn{1}{p{4cm}}{\raggedright TC3} & 
		\multicolumn{1}{p{4cm}}{\raggedright -1.560e-1*** (3.943e-2)} & 
		\multicolumn{1}{p{4cm}}{\raggedright -5.321e-2** (2.024e-2)} & 
		\multicolumn{1}{p{4cm}}{\raggedright -2.798e-2 (1.711e-2)} \\   
		\multicolumn{1}{p{4cm}}{\raggedright Age Purchased} & 
		\multicolumn{1}{p{4cm}}{\raggedright -6.723e-3**  (2.206e-3)} & 
		\multicolumn{1}{p{4cm}}{\raggedright -7.117e-3*** (8.395e-4)} & 
		\multicolumn{1}{p{4cm}}{\raggedright -5.208e-3*** (7.137e-4)} \\ 
		\multicolumn{1}{p{4cm}}{\raggedright (Age Purchased)$^2$} & 
		\multicolumn{1}{p{4cm}}{\raggedright 6.472e-5*** (1.899e-5)} & 
		\multicolumn{1}{p{4cm}}{\raggedright 5.841e-5*** (7.160e-6)} & 
		\multicolumn{1}{p{4cm}}{\raggedright 4.383e-5*** (5.999e-6)} \\ [1ex]
		\multicolumn{1}{p{4cm}}{\raggedright NZ Deprivation Score} & 
		\multicolumn{1}{p{4cm}}{\raggedright -8.780e-4*** (1.893e-4)} & 
		\multicolumn{1}{p{4cm}}{\raggedright -6.433e-4*** (8.427e-5)} & 
		\multicolumn{1}{p{4cm}}{\raggedright -6.206e-4*** (6.974e-5)} \\  
		\multicolumn{1}{p{4cm}}{\raggedright Burnside} & 
		\multicolumn{1}{p{4cm}}{\raggedright -8.311e-2 (2.126e-1)} & 
		\multicolumn{1}{p{4cm}}{\raggedright 9.399e-2*** (2.577e-2)} & 
		\multicolumn{1}{p{4cm}}{\raggedright 3.261e-2 (2.217e-2)} \\
		\multicolumn{1}{p{4cm}}{\raggedright BurnsideCHCBoys} & 
		\multicolumn{1}{p{4cm}}{\raggedright 4.097e-1$\dagger$ (2.116e-1) } & 
		\multicolumn{1}{p{4cm}}{\raggedright 2.914e-1*** (4.768e-2)} & 
		\multicolumn{1}{p{4cm}}{\raggedright 2.074e-1*** (3.975e-2)} \\
		\multicolumn{1}{p{4cm}}{\raggedright BurnsideCHCBoysGirls} & 
		\multicolumn{1}{p{4cm}}{\raggedright -} & 
		\multicolumn{1}{p{4cm}}{\raggedright -} & 
		\multicolumn{1}{p{4cm}}{\raggedright 11.976e-1 (1.536e-1)} \\
		\multicolumn{1}{p{4cm}}{\raggedright Cashmere} & 
		\multicolumn{1}{p{4cm}}{\raggedright -1.624e-1 (1.247e-1) } & 
		\multicolumn{1}{p{4cm}}{\raggedright 9.841e-2*** (2.210e-2)} & 
		\multicolumn{1}{p{4cm}}{\raggedright 6.181e-2** (1.973e-2 )} \\
		\multicolumn{1}{p{4cm}}{\raggedright CHCBoys} & 
		\multicolumn{1}{p{4cm}}{\raggedright -1.058e-2 (2.204e-1)} & 
		\multicolumn{1}{p{4cm}}{\raggedright 2.210e-1*** (5.415e-2)} & 
		\multicolumn{1}{p{4cm}}{\raggedright 1.356e-1** (4.247e-2)} \\
		\multicolumn{1}{p{4cm}}{\raggedright CHCGirlsBoys} & 
		\multicolumn{1}{p{4cm}}{\raggedright 6.225e-1*** (1.305e-1)} & 
		\multicolumn{1}{p{4cm}}{\raggedright 4.331e-1*** (4.265e-2} & 
		\multicolumn{1}{p{4cm}}{\raggedright  3.938e-1*** (3.582e-2)} \\
		\multicolumn{1}{p{4cm}}{\raggedright Riccarton} & 
		\multicolumn{1}{p{4cm}}{\raggedright -1.016e-3 (3.112e-1)} & 
		\multicolumn{1}{p{4cm}}{\raggedright 2.827e-2 (4.030e-2)} & 
		\multicolumn{1}{p{4cm}}{\raggedright -} \\
		\multicolumn{1}{p{4cm}}{\raggedright RiccartonCHCBoys} & 
		\multicolumn{1}{p{4cm}}{\raggedright -} & 
		\multicolumn{1}{p{4cm}}{\raggedright 3.320e-1* (1.574e-1)} & 
		\multicolumn{1}{p{4cm}}{\raggedright -} \\
		\multicolumn{1}{p{4cm}}{\raggedright RiccartonCHCGirlsBoys} & 
		\multicolumn{1}{p{4cm}}{\raggedright -} & 
		\multicolumn{1}{p{4cm}}{\raggedright -} & 
		\multicolumn{1}{p{4cm}}{\raggedright -} \\
		\multicolumn{1}{p{4cm}}{\raggedright Burwood} & 
		\multicolumn{1}{p{4cm}}{\raggedright -5.058e-2 (3.608e-2)} & 
		\multicolumn{1}{p{4cm}}{\raggedright -6.472e-2*** (1.654e-2)} & 
		\multicolumn{1}{p{4cm}}{\raggedright -7.210e-2*** (1.393e-2)} \\
		\multicolumn{1}{p{4cm}}{\raggedright Fendalton} & 
		\multicolumn{1}{p{4cm}}{\raggedright -} & 
		\multicolumn{1}{p{4cm}}{\raggedright 1.214e-1** (3.975e-2)} & 
		\multicolumn{1}{p{4cm}}{\raggedright 1.878e-1*** (3.385e-2)} \\ 
		\multicolumn{1}{p{4cm}}{\raggedright Linwood} & 
		\multicolumn{1}{p{4cm}}{\raggedright -1.591e-1 (1.101e-1) } & 
		\multicolumn{1}{p{4cm}}{\raggedright -1.034e-1*** (2.721e-2)} & 
		\multicolumn{1}{p{4cm}}{\raggedright -1.350e-1*** (2.182e-2)} \\ 
		\multicolumn{1}{p{4cm}}{\raggedright Papanui} & 
		\multicolumn{1}{p{4cm}}{\raggedright -2.557e-2 (1.010e-1) } & 
		\multicolumn{1}{p{4cm}}{\raggedright 5.253e-2** (1.795e-2)} & 
		\multicolumn{1}{p{4cm}}{\raggedright 5.607e-2*** (1.533e-2)} \\ 
		\multicolumn{1}{p{4cm}}{\raggedright Riccarton} & 
		\multicolumn{1}{p{4cm}}{\raggedright 2.063e-1 (3.678e-1)} & 
		\multicolumn{1}{p{4cm}}{\raggedright 7.751e-2* (3.064e-2)} & 
		\multicolumn{1}{p{4cm}}{\raggedright 1.066e-1*** (2.637e-2)} \\ 
		\multicolumn{1}{p{4cm}}{\raggedright Spreydon} & 
		\multicolumn{1}{p{4cm}}{\raggedright  -5.600e-2 (9.363e-2)}& 
		\multicolumn{1}{p{4cm}}{\raggedright 1.570e-2 (2.378e-2)} & 
		\multicolumn{1}{p{4cm}}{\raggedright  1.510e-2 (1.964e-2)} \\ 
		\multicolumn{1}{p{4cm}}{\raggedright Waimairi} & 
		\multicolumn{1}{p{4cm}}{\raggedright 1.125e-1 (2.313e-1) } & 
		\multicolumn{1}{p{4cm}}{\raggedright 5.472e-2  (2.774e-2 )} & 
		\multicolumn{1}{p{4cm}}{\raggedright 8.780e-2*** (2.377e-2)} \\ 
		\hline\hline \\[-1.5ex]
		\multicolumn{1}{p{4cm}}{\raggedright Degrees of Freedom} & 
		\multicolumn{1}{p{4cm}}{\raggedright 1054} & 
		\multicolumn{1}{p{4cm}}{\raggedright 1055} & 
		\multicolumn{1}{p{4cm}}{\raggedright 1462} \\
		\multicolumn{1}{p{4cm}}{\raggedright $R^2$} & 
		\multicolumn{1}{p{4cm}}{\raggedright 0.7099} & 
		\multicolumn{1}{p{4cm}}{\raggedright 0.8056} & 
		\multicolumn{1}{p{4cm}}{\raggedright 0.8216} \\ 
		\multicolumn{1}{p{4cm}}{\raggedright $R^2$ adjusted} & 
		\multicolumn{1}{p{4cm}}{\raggedright 0.6733} & 
		\multicolumn{1}{p{4cm}}{\raggedright 0.799} & 
		\multicolumn{1}{p{4cm}}{\raggedright 0.8172} \\ 
		\multicolumn{1}{p{4cm}}{\raggedright F-statistic} & 
		\multicolumn{1}{p{4cm}}{\raggedright 19.39} & 
		\multicolumn{1}{p{4cm}}{\raggedright 121.3} & 
		\multicolumn{1}{p{4cm}}{\raggedright 187.1} \\ 
		\hline
		\caption{Note. Table of coefficients for the quarterly time dummy model in different time periods. The legends '***' indicate significance at a 0.001 level, '**' indicates significance at a 0.01 level, '*' indicates significance at a 0.05 level, and '$\dagger$' indicates significance at a 0.1 level. }
	\end{longtable}
	
	\section{Suburbs and Technical Categories}\label{app:additional}
	
	Table \ref{tab:suburb_results} contains the summary results for the OLS models for each period when suburbs are used instead of wards as the geographic feature of each property. As mentioned in Section \ref{sec:method}, the use of suburbs better specifies the model; however, the collinearity of the suburbs with the technical category of the property makes it unsuitable to be used in conjunction with TC which is the variable we are interested in.

	\begin{table}[!ht]
		\centering
		\begin{tabular}{l l l l}
			\hline\hline \\ 
			~ & \textbf{2014-2016} & \textbf{2019-2020} & \textbf{2020-2021} \\
			\hline
			Degrees of Freedom & 267 & 1001 & 1407\\
			R-squared & 0.8378 & 0.8389 & 0.8594\\ 
			Adj R-squared & 0.7886 & 0.8246 & 0.8503\\ 
			F-Statistic & 17.03 & 58.56 & 94.51\\
			\hline
		\end{tabular}
		\caption{Summary results for OLS models using suburb in place of wards as the geographical feature.}
		\label{tab:suburb_results}

	\end{table}

	Table \ref{tab:contigency} shows the contingency table for each period for the properties' suburb and technical category. This table gives the view that if suburbs were to be used in place of wards as the geographical identifiers it will severely overlap with technical category in terms of the information that it retrieves, specifically in the case of TC1.
	
	\setlength{\LTleft}{-20cm plus -1fill}
	\setlength{\LTright}{\LTleft}
	
	\begin{longtable}{l|lll|lll|lll}
		\hline\hline
		\textbf{} & \multicolumn{3}{c|}{\textbf{2014-2016}} & \multicolumn{3}{c|}{\textbf{2019-2020}}  & \multicolumn{3}{c}{\textbf{2020-2021}}  \\ 
		\textbf{Suburb} & TC1 & TC2 & TC3 & TC1 & TC2 & TC3 & TC1 & TC2 & TC3 \\ \hline
		\textbf{Aidanfield} & 0 & 2 & 0 & 0 & 9 & 0 & 0 & 18 & 0 \\ 
		\textbf{Aranui} & 0 & 2 & 3 & 0 & 11 & 10 & 0 & 15 & 14 \\ 
		\textbf{Avondale} & 0 & 3 & 2 & 0 & 1 & 13 & 0 & 4 & 22 \\ 
		\textbf{Avonhead} & 13 & 0 & 0 & 49 & 0 & 0 & 66 & 0 & 0 \\
		\textbf{Avonside} & 0 & 3 & 3 & 0 & 6 & 3 & 0 & 5 & 3 \\
		\textbf{Beckenham} & 0 & 1 & 0 & 0 & 8 & 0 & 0 & 9 & 2 \\ 
		\textbf{Belfast} & 0 & 3 & 0 & 0 & 8 & 0 & 0 & 7 & 1 \\ 
		\textbf{Bishopdale} & 2 & 13 & 1 & 8 & 25 & 1 & 16 & 37 & 3 \\
		\textbf{Bromley} & 0 & 2 & 0 & 0 & 9 & 0 & 0 & 13 & 0 \\ 
		\textbf{Broomfield} & 6 & 0 & 0 & 8 & 0 & 0 & 14 & 0 & 0 \\
		\textbf{Bryndwr} & 2 & 7 & 0 & 3 & 25 & 0 & 6 & 37 & 0 \\ 
		\textbf{Burnside} & 13 & 1 & 0 & 34 & 10 & 1 & 58 & 7 & 1 \\
		\textbf{Burwood} & 0 & 6 & 7 & 0 & 25 & 26 & 0 & 27 & 36 \\ 
		\textbf{Casebrook} & 0 & 7 & 1 & 0 & 23 & 3 & 0 & 34 & 5 \\ 
		\textbf{Cashmere} & 0 & 2 & 1 & 0 & 9 & 3 & 0 & 6 & 0 \\ 
		\textbf{Cracroft} & ~ & ~ & ~ & 0 & 1 & 0 & 0 & 1 & 0 \\ 
		\textbf{Dallington} & 0 & 2 & 5 & 0 & 5 & 4 & 0 & 10 & 10 \\ 
		\textbf{Fendalton} & 0 & 6 & 1 & 0 & 10 & 10 & 0 & 30 & 6 \\ 
		\textbf{Halswell} & 0 & 15 & 2 & 0 & 64 & 6 & 0 & 70 & 6 \\ 
		\textbf{Harewood} & ~ & ~ & ~ & 2 & 13 & 0 & 0 & 14 & 0 \\ 
		\textbf{Heathcote Valley} & 0 & 2 & 0 & 0 & 8 & 0 & 0 & 6 & 0 \\ 
		\textbf{Hei Hei} & 6 & 0 & 0 & 17 & 0 & 0 & 16 & 0 & 0 \\ 
		\textbf{Hillmorton} & 0 & 6 & 1 & 0 & 13 & 3 & 0 & 21 & 3 \\ 
		\textbf{Hillsborough} & 0 & 2 & 0 & 0 & 5 & 1 & 0 & 8 & 1 \\ 
		\textbf{Hoon Hay} & 0 & 20 & 0 & 0 & 33 & 0 & 0 & 50 & 2 \\ 
		\textbf{Hornby} & 14 & 0 & 0 & 30 & 0 & 0 & 38 & 0 & 0 \\ 
		\textbf{Huntsbury} & ~ & ~ & ~ & ~ & ~ & ~ & 0 & 1 & 0 \\ 
		\textbf{Ilam} & 7 & 4 & 0 & 17 & 11 & 0 & 33 & 20 & 0 \\ 
		\textbf{Islington} & 1 & 0 & 0 & 6 & 0 & 0 & 9 & 0 & 0 \\ 
		\textbf{Kainga} & ~ & ~ & ~ & ~ & ~ & ~ & 0 & 1 & 0 \\ 
		\textbf{Linwood} & 0 & 3 & 2 & 0 & 6 & 4 & 0 & 11 & 5 \\ 
		\textbf{Mairehau} & 0 & 8 & 1 & 0 & 19 & 8 & 0 & 30 & 9 \\ 
		\textbf{Merivale} & 0 & 0 & 4 & 0 & 3 & 11 & 0 & 2 & 14 \\ 
		\textbf{Mount Pleasant} & ~ & ~ & ~ & 0 & 3 & 0 & 0 & 3 & 0 \\ 
		\textbf{New Brighton} & 0 & 3 & 3 & 0 & 8 & 19 & 0 & 19 & 24 \\ 
		\textbf{North New Brighton} & 0 & 4 & 2 & 0 & 7 & 8 & 0 & 14 & 11 \\ 
		\textbf{Northcote} & 0 & 3 & 0 & 0 & 9 & 0 & 0 & 11 & 0 \\ 
		\textbf{Northwood} & 0 & 2 & 2 & 0 & 12 & 4 & 0 & 15 & 9 \\ 
		\textbf{Opawa} & 0 & 0 & 2 & 0 & 1 & 6 & 0 & 4 & 7 \\ 
		\textbf{Papanui} & 0 & 7 & 0 & 0 & 17 & 0 & 0 & 38 & 0 \\ 
		\textbf{Parklands} & 0 & 4 & 9 & 0 & 23 & 43 & 0 & 28 & 44 \\ 
		\textbf{Redcliffs} & 0 & 1 & 0 & 0 & 2 & 1 & 0 & 3 & 2 \\ 
		\textbf{Redwood} & 0 & 16 & 0 & 0 & 54 & 1 & 0 & 53 & 1 \\ 
		\textbf{Riccarton} & ~ & ~ & ~ & 0 & 7 & 0 & 0 & 6 & 0 \\ 
		\textbf{Richmond} & 0 & 0 & 3 & 0 & 2 & 8 & 0 & 2 & 13 \\ 
		\textbf{Russley} & 5 & 0 & 0 & 14 & 0 & 0 & 15 & 0 & 0 \\ 
		\textbf{Saint Martins} & 0 & 3 & 1 & 0 & 6 & 6 & 0 & 11 & 8 \\ 
		\textbf{Shirley} & 0 & 11 & 2 & 0 & 18 & 2 & 0 & 31 & 3 \\ 
		\textbf{Sockburn} & 2 & 0 & 0 & 18 & 0 & 0 & 18 & 0 & 0 \\ 
		\textbf{Somerfield} & 0 & 2 & 0 & 0 & 19 & 0 & 0 & 9 & 0 \\ 
		\textbf{South New Brighton} & 0 & 2 & 0 & 0 & 7 & 2 & 0 & 11 & 3 \\ 
		\textbf{Southshore} & 0 & 0 & 1 & 0 & 4 & 6 & 0 & 3 & 2 \\ 
		\textbf{Spreydon} & 0 & 2 & 0 & 0 & 6 & 0 & 0 & 3 & 0 \\ 
		\textbf{St Albans} & 0 & 4 & 6 & 0 & 7 & 10 & 0 & 20 & 20 \\ 
		\textbf{Strowan} & 0 & 6 & 0 & 0 & 9 & 1 & 0 & 23 & 3 \\ 
		\textbf{Sumner} & 0 & 4 & 0 & 0 & 4 & 0 & 0 & 10 & 0 \\ 
		\textbf{Sydenham} & ~ & ~ & ~ & 0 & 3 & 0 & 0 & 3 & 0 \\ 
		\textbf{Templeton} & ~ & ~ & ~ & 4 & 0 & 0 & 12 & 0 & 0 \\ 
		\textbf{Upper Riccarton} & 3 & 1 & 0 & 8 & 6 & 0 & 6 & 8 & 0 \\ 
		\textbf{Waimairi Beach} & 0 & 2 & 0 & 0 & 8 & 0 & 0 & 7 & 0 \\ 
		\textbf{Wainoni} & 0 & 0 & 5 & 0 & 2 & 8 & 0 & 6 & 9 \\ 
		\textbf{Westmorland} & ~ & ~ & ~ & ~ & ~ & ~ & 0 & 1 & 0 \\ 
		\textbf{Wigram} & 1 & 0 & 0 & 8 & 0 & 0 & 15 & 0 & 0 \\ 
		\textbf{Woolston} & 0 & 3 & 4 & 0 & 16 & 8 & 0 & 25 & 12 \\ 
		\textbf{Yaldhurst} & ~ & ~ & ~ & 2 & 0 & 0 & 2 & 0 & 0 \\ 
		\textbf{Moncks Bay} & ~ & ~ & ~ & 0 & 3 & 0 & ~ & ~ & ~ \\ 
		\caption{Contigency table for suburb and technical category for each time period.}
		\label{tab:contigency}

	\end{longtable}

	\end{document}